\documentclass[a4paper,11pt]{article}

\usepackage{jheppub}

\title{Axion identification in multi-phase models}

\author[a]{Zhen-Yu Lei}
\author[a,1]{Jian-Wei Cui\note{Corresponding author.}}

\affiliation[a]{School of Physics, Huazhong University of Science and Technology, Wuhan 430074, China.}

\emailAdd{jwcui@hust.edu.cn}

\abstract{
	Axion identification in multi-phase models is subtle because local mass eigenstates do not by themselves determine global properties such as the fundamental period and domain-wall number.
	In this paper, we develop a route to identify the axion in a broad class of multi-phase models, including DFSZ-like theories. The construction involves two reductions: gauge identification and integrating out heavy phase fields. We study these reductions explicitly in two representative two-phase models. Gauge identification reduces the compact field space of the phase fields to a quotient space by identifying points connected by gauge transformations, while integrating out heavy phase fields restricts the low-energy dynamics to nonlinear orbits. Under the stated assumptions and within the validity of the derivative expansion, the physical low-energy field space forms a finite union of $S^1$. The axion is identified as the effective field on an $S^1$ orbit and coincides locally with the lightest mass eigenstate near the vacuum. Its fundamental period and domain-wall number are well-defined and can be determined from the original Lagrangian. Nevertheless, the reduction to a single axion does not capture all properties of topological defects in the original multi-phase theory. In particular, axion strings can retain nontrivial gauge flux, while special defect configurations can involve non-negligible spacetime derivatives of the heavy phase fields and invalidate the single-field reduction. These effects represent fundamental differences between multi-phase and single-phase theories at the level of topological defects.

}

\begin{document}
	
	\maketitle
	\flushbottom
	
	\section{Introduction}
	\label{sec:intro}
	The axion, the pseudo-Goldstone boson (pNGB) associated with the breaking of the Peccei–Quinn (PQ) symmetry~\cite{PQ1,PQ2}, was originally introduced as a solution to the strong CP problem~\cite{PhysRevLett.40.223,PhysRevLett.40.279,Kim-KSVZ,SHIFMAN1980493,DINE1981199,Zhitnitskii1980}. In addition to being a compelling dark-matter candidate \cite{Preskill1983,Abbott1983,Dine1983,TURNER199067,kim2010review}, the axion has rich implications for cosmology. If PQ symmetry breaking occurs after inflation, the vacuum expectation value (VEV) of the axion field can take different values in causally disconnected regions of space \cite{Kibble1976,Kibble1980,Murayama2010}. Spatial variations of this VEV then give rise to topological defects in the form of cosmic strings \cite{Nielsen1973,Dvali1993,Vilenkin1985,Abe2021,Eto2022,Weinberg2012}. After the explicit breaking turns on, these axion strings become attached to domain walls (DWs) \cite{Sikivie1982,Di2020}, forming string–wall networks \cite{Vilenkin1982,Kibble1982,Chang1998,Dunsky2024,Dvali2025,Li2023}.	
	
	In standard single-phase theories, the axion is defined as a periodic scalar field \cite{Agrawal2018,Ernst2018,Fraser2020,Agrawal2022,Diehl2023,Di2020,Marsh2016}
	\begin{equation}
		a(x)\cong a(x)+2\pi F_a,
	\end{equation}
	with the standard cosine potential 
	\begin{equation}
		V(a)=\Lambda^4[1-\cos(N_{DW}a/F_a)],\label{intro:Va}
	\end{equation}
	where $2\pi F_a$ is the fundamental period of the axion, $\Lambda$ represents the scale of explicit symmetry breaking, and $N_{DW}$ is a positive integer. The axion decay constant is defined by $f_a\equiv F_a/N_{DW}$. This periodicity indicates that the field space of the axion, called the axion orbit in this paper, is homeomorphic to $S^1$. This topology leads to the existence of axion strings and gives $N_{DW}$ its physical meaning: the number of vacua within a single period, which equals the number of DWs attached to a fundamental axion string.
	
	However, mixing between phases makes axion identification in multi-phase models less straightforward. One conventional method identifies the axion as the Goldstone mode of PQ symmetry \cite{Srednicki1985,Ernst2018,Di2020,Sun2020,Diehl2023}
	\begin{equation}
		a=\frac{1}{f_{PQ}}\sum_iq_iv_i^2\theta_i.
	\end{equation}
	where $q_i,v_i/\sqrt{2}$ are the PQ charges and VEVs of the PQ fields, respectively, while $f_{PQ}^2\equiv\sum_iq_i^2v_i^2$. The $\theta_i$ are the dimensionless phases of the PQ fields, each with period $2\pi$. A subtlety of this identification is that, in the presence of additional gauge symmetries, the PQ generator can be redefined by adding other symmetry generators, so the PQ charges need not be uniquely determined even after fixing their overall normalization. One way to address this ambiguity is to impose orthogonality between the Goldstone modes associated with PQ symmetry and those associated with gauge symmetries \cite{DINE1981199}. This procedure may lead to irrational PQ charges, raising concerns about the compactness of $U(1)_{PQ}$ and whether the axion period and $N_{DW}$ are well-defined \cite{Ernst2018,Buen2021,Reece2023,Diehl2023,Choi2026}. The compactness of the low-energy axion in the presence of mixing was discussed in Ref.~\cite{Fraser2020}. There, it was shown that an exactly massless axion inherits the periodicity and the associated constraints of the compact UV theory. When the axion is massive, Ref.~\cite{Fraser2020} instead focused on how these constraints can be relaxed. A general conclusion on whether the massive low-energy axion itself remains periodic or compact was not established there.
	
	In this paper, we develop a method for identifying the axion in a broad class of axion models that satisfy the following main assumptions:
	\begin{itemize}
		\item[1.] The models contain multiple scalar fields $S_i$ that can be parameterized nonlinearly as
		\begin{equation}
			S_i=\frac{1}{\sqrt{2}}(\rho_i+f_i)\mathrm{e}^{\mathrm{i}\phi_i/f_i},
		\end{equation}
		where $f_i/\sqrt{2}$ is the VEV of $S_i$ and is called the spontaneous breaking scale in this paper. The radial field $\rho_i$ is not relevant to the present discussion. We usually express the phase field $\phi_i$ in terms of the dimensionless field $\theta_i\equiv\phi_i/f_i$ with period $2\pi$.
		\item[2.] In the mass spectrum of these phase fields, all massless fields are eaten by gauge fields. Among the remaining massive fields, a mass gap separates the single lightest phase field from the other heavy phase fields.
	\end{itemize} 
	In addition to DFSZ-like models \cite{Darme2020,Cox2024,Babu2024,Rocha2025,Babu2026,Giraldo2026}, extended KSVZ models \cite{Lei2024} and multi-axion models \cite{Kim2005,Mack2011,Choi2014,Bachlechner2018,Murai2023,Lee2024,Li2025,Lee2025,Lee2026} can satisfy the above assumptions. The field-content assumption can also be relaxed by extending the phase-only description to include other periodic pseudoscalar fields, such as the pion, $\eta$, and $\eta'$ \cite{Dvali2025}. Beyond these main assumptions, we make one further assumption to simplify the calculations: 
	\begin{itemize}
		\item[3.] The scalar potential in the N-phase model is
		\begin{equation}
			V(\theta_1,\dots,\theta_N)=\sum_{i=1}^n \Lambda_i^4(1-\cos X_i),\label{intro:Vtheta}
		\end{equation}
		where $X_1,\dots,X_n$ are linearly independent integer-coefficient combinations of $\theta_i$, and $n$ is the number of massive phase fields. $\Lambda_i$ is called the explicit breaking scale in this paper.
	\end{itemize}
	Although this constraint on the potential limits the applicability of the method to some extent, it significantly simplifies the calculations.\footnote{For example, the constant in the cosine, e.g. $\cos(X_i+c_i)$, can always be absorbed through redefinitions of the $X_i$, and the vacua of the potential are always located at $X_i=2\pi n_i$.}
	Our discussion is not restricted to QCD axion models. For QCD axion models, the scalar potential of these massive phases involves the QCD instanton induced term to solve the strong CP problem. 	
	
	In this paper, we identify the axion using a method that does not require an explicit assignment of PQ charges. The identification involves two key reduction mechanisms: (i) gauge identification and (ii) integrating out heavy fields. Provided that the derivative expansion of the equation of motions (EOMs) of heavy fields is valid, the physical reduced field space is homeomorphic to a finite union of $S^1$. On each $S^1$, the field period and $N_{DW}$ are both well-defined. The axion is identified as the effective field on a selected $S^1$ and coincides with the lightest phase field near the vacuum.
	
	To understand the two reduction mechanisms, we first consider two benchmark two-phase models: (i) a light phase field mixed with a gauged phase field that is eaten by a gauge field, and (ii) a light phase field mixed with a heavier phase field. In both cases, mixing generically makes the orbit of the lightest phase non-compact. Identifying the axion solely with the lightest phase would therefore yield an axion without a period. In the first model, however, gauge identification allows us to factor out the gauge redundancies. The quotient field space is homeomorphic to $S^1$. The axion is defined on this quotient field space and has a finite period and a well-defined $N_{DW}$. In the second model, we can integrate out the heavy field using its EOM. If the spacetime dependence in the EOM can be neglected, the EOM becomes an algebraic constraint that restricts the field space from $T^2$ to a union of $S^1$. The axion is identified as the effective field on one of these $S^1$ orbits, with a finite period and a well-defined $N_{DW}$. In this sense, our analysis extends the compact low-energy description established for an exactly massless axion in Ref.~\cite{Fraser2020} to the massive case under the assumptions specified above. 
	
	Combining these two reduction mechanisms, we obtain a route to axion identification in multi-phase models: 
	\begin{equation*}
		T^N\xrightarrow{\text{gauge identification}}T^{N-k}\xrightarrow{\text{integrate out heavy fields}}\bigcup_i S^1_i.
	\end{equation*}
	The axion is identified as the effective field on one of these $S^1$ orbits, and the corresponding period and $N_{DW}$ can both be computed from the original Lagrangian.

	However, the above reductions may encounter two difficulties when applied to nontrivial spacetime-dependent configurations, particularly topological defects. On the one hand, the gauge identification in the first reduction mechanism is pointwise and therefore does not retain the information carried by the gauge field, such as the gauge flux of an axion string. 
	On the other hand, for some defect configurations, such as the string–wall networks involving heavy phase fields discussed in Ref.~\cite{Lee2025}, spatial derivatives of the heavy fields cannot be neglected within the domain walls, invalidating the  reduction used in the second mechanism.
	These phenomena reveal fundamental differences between multi-phase and single-phase models at the level of topological defects, as also discussed in Refs.~\cite{Lee2025,Choi2026}.
	
	The remainder of this paper is organized as follows. In section~\ref{sec:eaten}, we discuss a simplified two-phase model where the massless phase field is eaten by the gauge field. In section~\ref{sec:heavy}, we discuss another simplified two-phase model where one phase field is much heavier than the other massive field. In section~\ref{sec:compact}, we generalize the results from the two-phase models with a concise proof and then combine them to obtain a route to axion identification. Finally, section~\ref{sec:conc} summarizes our main results.
	
	\section{Mixing with the gauged phase field}
	\label{sec:eaten}
	In this section, we consider a simplified two-phase model containing two phase fields $\phi_1$ and $\phi_2$ with fundamental periods
	\begin{equation}
		\phi_{1,2}\cong\phi_{1,2}+2\pi f_{1,2}.\label{eaten:phi12}
	\end{equation}
	To normalize the periods to $2\pi$, we can define the following dimensionless phase fields
	\begin{equation}
		\theta_{1,2}=\frac{\phi_{1,2}}{f_{1,2}}.\label{eaten:thetaphi}
	\end{equation}
	The field space of $(\theta_1,\theta_2)$ is a 2-dimensional torus $T^2$. The winding numbers on this torus can be represented by $(k_1,k_2)$, where $k_1,k_2$ are the winding numbers of $\theta_1,\theta_2$, respectively.
	
	\subsection{The mass eigenstates and $a$-orbit}
	The relevant Lagrangian in this model is 
	\begin{equation}
		\begin{aligned}
			\mathcal{L}\supset\sum_{i=1}^{2}\frac{f_i^2}{2}(\partial_\mu\theta_i-gq_iA_\mu)^2-V(\theta_1,\theta_2),
		\end{aligned}
	\end{equation}
	where $q_1$ and $q_2$ are the charges under the $U(1)$ symmetry gauged by $A_\mu$, with gauge transformations
	\begin{equation}
		A_\mu\rightarrow A_\mu+\partial_\mu\alpha,\quad \theta_1\rightarrow\theta_1+q_1g\alpha,\quad \theta_2\rightarrow\theta_2+q_2g\alpha,\label{eaten:U(1)}
	\end{equation}
	and $g$ is the corresponding gauge coupling. The potential takes the form 
	\begin{equation}
		V(\theta_1,\theta_2)=\Lambda^4\;[1-\cos(n_1\theta_1+n_2\theta_2)],\label{eaten:Va}
	\end{equation}
	which is the special case of Eq.~\eqref{intro:Vtheta} with $N=2$ and $n=1$. Here, the coefficients $n_1$ and $n_2$ are both integers.
	If one of $n_1$ and $n_2$ vanishes, then $(\theta_1,\theta_2)$ themselves are mass eigenstates, and the gauge identification simply amounts to setting $\theta_1=0$ or $\theta_2=0$. Then the axion is globally identified as either $\theta_2$ or $\theta_1$. Here, we consider the more complicated case in which the mass eigenstates are mixtures of $(\theta_1,\theta_2)$, i.e., both $n_1$ and $n_2$ are nonzero.
	
	Invariance under the $U(1)$ gauge transformation in Eq.~\eqref{eaten:U(1)} requires
	\begin{equation}
		n_1q_1+n_2q_2=0.\label{eaten:nkq}
	\end{equation}
	This implies that the ratio of the charges $q_1/q_2$ is rational. Thus, we can always rescale all the gauge charges to integers, obtaining a well-defined group action.
	
	We can diagonalize the mass term using the following transformation
	\begin{equation}
		\begin{pmatrix}
			\theta_1\\\theta_2
		\end{pmatrix}=f_a\begin{pmatrix}
			\frac{n_1}{f_1^2}&-\frac{n_2}{f_1f_2}\\
			\frac{n_2}{f_2^2}&\frac{n_1}{f_1f_2}
		\end{pmatrix}\begin{pmatrix}
			a\\b
		\end{pmatrix}.\label{eaten:thetaa}
	\end{equation}
	where 
	\begin{equation}
		f_a=\frac{f_1f_2}{\sqrt{n_1^2 f_2^2+n_2^2 f_1^2}}
	\end{equation}
	is the axion decay constant, as shown in Eq.~\eqref{eaten:L}. Notably, this transformation is not orthogonal.
	Rewriting the Lagrangian in terms of $a$ and $b$ gives
	\begin{equation}
		\mathcal{L}\supset\frac{1}{2}\partial_\mu a\partial^\mu a+\frac{1}{2}\big(\partial_\mu b+\mu_AA_\mu \big)^2-\Lambda^4\;[1-\cos(a/f_a)],\label{eaten:L}
	\end{equation}
	where
	\begin{equation}
		\mu_A=\frac{gq_1}{n_2}\sqrt{n_1^2f_2^2+n_2^2f_1^2}.
	\end{equation}
	It can be verified that $b$ is the Goldstone mode eaten by $A_\mu$. The masses of $A_\mu$ and $a$ can be read off from the Lagrangian \eqref{eaten:L}
	\begin{equation}
		m_A^2=\mu_A^2,\quad m_a^2=\frac{\Lambda^4}{f_a^2}.
	\end{equation}
	In this paper, we refer to a branch of the field space of $a$, i.e., the trajectory obtained by varying $a$ while keeping $b$ fixed, as the $a$-orbit. Similarly, we refer to a branch of the field space of $b$ as the $b$-orbit. Because the $b$-orbit is generated by gauge transformations, it is also called the gauge orbit \cite{Achcarro2000}.

	The explicit form of the $a$-orbit follows from the relation~\eqref{eaten:thetaa}, with the parametrization (at fixed $b=0$)
	\begin{equation}
		\theta_1=\frac{n_1f_a}{f_1^2}a,\quad \theta_2=\frac{n_2f_a}{f_2^2}a.
	\end{equation}
	Thus, the periodicity of the $a$-orbit depends on its slope\footnote{By contrast, one can verify that the slope of the gauge orbit is always rational because the gauge charges are integers.}
	\begin{equation}
		s_a=\frac{n_2}{n_1}\frac{f_1^2}{f_2^2}.\label{eaten:slope}
	\end{equation}
	\begin{itemize}
		\item $s_a$ is rational\\
		In this scenario, we can always find an integer pair $(m_1,m_2)$ satisfying
		\begin{equation}
			\frac{m_2}{m_1}=s_a,\quad \gcd(m_1,m_2)=1,
		\end{equation}
		where $\gcd(m_1,m_2)$ is the greatest common divisor of $m_1$ and $m_2$. Then the $a$-orbit is homeomorphic to $S^1$ with the fundamental period
		\begin{equation}
			a\cong a+2\pi F_a,\quad F_a=\Big|\frac{m_1}{n_1}\Big|\frac{f_1^2}{f_a}.
		\end{equation}
		Here, string configurations exist along this orbit with a finite DW number
		\begin{equation}
			N^{(a\text{-orbit})}_{DW}=|n_1m_1+n_2m_2|.\label{eaten:NDW1}
		\end{equation}
		\item $s_a$ is irrational\\
		In this scenario, the $a$-orbit is a non-closed, dense curve.
		
		This implies that the length of the $a$-orbit is infinite and no fundamental period exists. Therefore, there is no string configuration along this orbit and the corresponding DW number is ill-defined \cite{Diehl2023}.
	\end{itemize}
	If we identify the Goldstone mode of $U(1)_{PQ}$ with $a$, the ratio of the PQ charges is exactly $s_a$. If $s_a$ is irrational, we cannot simultaneously rescale the PQ charges to integers. This indicates that the group action of $U(1)_{PQ}$ is ill-defined, i.e., the same group element corresponds to different actions. Equivalently, the PQ symmetry is not a $U(1)$ symmetry but an $\mathbb{R}$ symmetry.  
	In the presence of mixing between $\theta_1$ and $\theta_2$, whether $s_a$ is rational or irrational depends on the ratio $f_1^2/f_2^2$. Without additional constraints, such as imposing an additional $\mathbb{Z}_2$ symmetry, this ratio is generically irrational. 
	
	Before proceeding, we emphasize that the periodicity of a field depends only on its field space. In particular, for $a$, the periodicity is determined by the slope $s_a$ in Eq.~\eqref{eaten:slope}. However, since the transformation between $a,b$ and $\theta_1,\theta_2$ in Eq.~\eqref{eaten:thetaa} is not orthogonal, the ratio of the coefficients of $\theta_2$ and $\theta_1$ in the expansion of $a$ 
	\begin{equation}
		a=f_a(n_1\theta_1+n_2\theta_2),\label{eaten:afa}
	\end{equation}
	is always rational, which generically differs from $s_a$. 
	
	\subsection{The axion orbit}\label{sec:aorbit}
	So far, we have obtained the mass eigenstates $a$ and $b$, and verified that the $a$-orbit is generically non-closed. However, as we will show below, the non-compact $a$-orbit can be made compact through gauge identification.
	
	The gauge orbit can be read off from the gauge transformation~\eqref{eaten:U(1)} or, equivalently, from the relation~\eqref{eaten:thetaa} (at fixed $a=0$)
	\begin{equation}
		\theta_1=-\frac{n_2f_a}{f_1f_2}b,\quad \theta_2=\frac{n_1f_a}{f_1f_2}b.
	\end{equation}
	If we can identify all points on the gauge orbit with a single point, the field space reduces to a quotient space, which we identify as the axion orbit.
	
	Because of the nontrivial topology of the torus, the gauge orbit intersects the $a$-orbit at points other than the origin. The condition for the two orbits to intersect is
	\begin{equation}
		\frac{n_1f_a}{f_1^2}a=-\frac{n_2f_a}{f_1f_2}b+2\pi m_1',\quad \frac{n_2f_a}{f_2^2}a=\frac{n_1f_a}{f_1f_2}b+2\pi m_2',
	\end{equation}
	where $m_1'$ and $m_2'$ are arbitrary integers.
	We then obtain
	\begin{equation}
		a=2\pi f_a(n_1m_1'+n_2m_2'),\quad b=2\pi f_a\frac{n_2m_1'f_1^2-n_1m_2'f_2^2}{f_1f_2}.\label{eaten:ab}
	\end{equation}
	The following theorem from number theory holds \cite{Hardy1979}: for a fixed integer pair $(n_1,n_2)$, 
	\begin{itemize}
		\item for any integer pair $(m_1',m_2')$, $(n_1m_1'+n_2m_2')$ is a multiple of $\operatorname{gcd}(n_1,n_2)$;
		\item there exist Bézout coefficients $(m_1',m_2')$ satisfying
		\begin{equation}
			n_1m_1'+n_2m_2'=\operatorname{gcd}(n_1,n_2).
		\end{equation}
	\end{itemize}
	For a given pair of Bézout coefficients $(m_1',m_2')$, the next intersection point after the origin along the $a$-orbit is $a=2\pi F_a$, with
	\begin{equation}
		F_a=f_a\operatorname{gcd}(n_1,n_2).\label{eaten:Fafa}
	\end{equation}
	Between $a=0$ and $a=2\pi F_a$, all points are gauge inequivalent. This implies that the trajectory from $a$ to $a+2\pi F_a$, with the identification
	\begin{equation}
		a\cong a+2\pi F_a,
	\end{equation}
	is the axion orbit. Thus, the axion orbit is homeomorphic to $S^1$ with period $2\pi F_a$. The DW number on the axion orbit can be obtained from Eq.~\eqref{eaten:Fafa}
	\begin{equation}
		N_{DW}=\operatorname{gcd}(n_1,n_2).\label{eaten:NDW}
	\end{equation}
	Although the axion can be viewed as $a$ locally, the contribution from $b$ cannot be omitted when considering global properties. It is precisely the contribution from $b$ that endows the axion with period $2\pi F_a$. However, this contribution is obscured once gauge identification is imposed. This is because the gauge identification used above is pointwise and is generically not equivalent to gauge fixing when the spacetime dependence of the fields is considered. The distinction between pointwise gauge identification and gauge fixing is discussed in section~\ref{sec:string}.
	
	Here, the axion identification shares a common feature with that in Ref.~\cite{Choi2026}: in both cases, we distinguish the Goldstone mode of the PQ symmetry from the mass eigenstates. However, in this paper the axion is identified in the quotient space, whereas in Ref.~\cite{Choi2026} the mass eigenstates are directly defined as the axion. We can pull back the axion orbit to the torus. The axion orbit is the quotient space of the torus under gauge equivalence, with the quotient map
	\begin{equation}
		\pi: T^2\rightarrow S^1,\quad (\theta_1,\theta_2)\mapsto [(\theta_1,\theta_2)],
	\end{equation}
	where $[(\theta_1,\theta_2)]$ is the gauge equivalence class of $(\theta_1,\theta_2)$. In the quotient space, the distance between two points is simply the minimum distance between their representatives on the torus, i.e., the distance along the $a$-orbit. We can define a continuous section of the quotient space
	\begin{equation}
		s:S^1\rightarrow T^2,\quad  \theta\mapsto(\theta_1,\theta_2).
	\end{equation}
	where $(\theta_1,\theta_2)$ is an element of the gauge equivalence class $\theta$.
	The Goldstone mode of the PQ symmetry can be regarded as the length along the section, where the length should be defined consistently with the quotient space rather than the torus. If we require that this section is tangent to the curve $b=0$ at the vacuum, the Goldstone mode coincides with $a$ near the vacuum. Notably, the section condition $\pi\circ s=\mathrm{id}$ guarantees that the section is a non-contractible loop in the torus.

	\subsection{Clarification of the DW number}
	In the preceding discussion, we obtained two different DW numbers for the $a$-orbit (if $s_a$ is rational) and the axion orbit. In Refs.~\cite{Ernst2018,Diehl2023,Babu2024}, the authors choose the mass eigenstate $a$ as the axion and calculate $N_{DW}$ as
	\begin{equation}
		N_{DW}=\text{minimal positive integer }\Big\{\frac{a(\theta_{i})|_{\theta_i=2\pi k_i}}{2\pi f_a},\; k_i\in\mathbb{Z}\Big\}.\label{eaten:NDW2}
	\end{equation}
	Using the expression for $a$ in Eq.~\eqref{eaten:afa}, we obtain
	\begin{equation}
		N_{DW}=\text{minimal positive integer }\{n_1k_1+n_2k_2\}=\operatorname{gcd}(n_1,n_2),
	\end{equation}
	which is consistent with Eq.~\eqref{eaten:NDW}. This is not a coincidence, since the right-hand side (RHS) of Eq.~\eqref{eaten:NDW2} can be understood as the change in $a$, divided by $2\pi f_a$, after traversing any loop with winding number $(k_1,k_2)$ that starts and ends at the origin. We can construct this loop as follows: starting at the origin, follow the $a$-orbit to one of the intersection points in Eq.~\eqref{eaten:ab}, then return to the origin along the $b$-orbit. Thus, the minimal change in $a$ during this process is exactly $2\pi F_a$. 
	
	This explanation gives the physical meaning of $N_{DW}$ in Eq.~\eqref{eaten:NDW2}: it is the number of vacua within a single period of the axion orbit. The compactness of the axion orbit guarantees that $N_{DW}$ exists and is well-defined.
	This also explains why $N^{(a\text{-orbit})}_{DW}$ for the $a$-orbit (if $s_a$ is rational) cannot be less than $N_{DW}$ for the axion orbit: if the $a$-orbit is closed, the intersection points divide it into equal segments, each corresponding to an axion orbit. This is exactly the Lazarides–Shafi (LS) mechanism \cite{Lazarides1982,Chatterjee2020}.

	\subsection{The axion string}\label{sec:string}
	After pointwise gauge identification, all contributions from $b$ are eliminated. It seems that this two-phase model reduces to a single-phase model with the standard potential \eqref{intro:Va}, where $F_a$ and $N_{DW}$ are given in Eqs.~\eqref{eaten:Fafa} and \eqref{eaten:NDW}. In this subsection, we show that this reduction loses physical information carried by the gauge field when an axion string is considered.
	
	We first consider the axion string alone, omitting the potential in Eq.~\eqref{eaten:Va}. Before gauge identification, the $a$-orbit is generically non-closed, and the so-called axion string is a mixture of the light phase field $a$ and the Goldstone mode $b$. Using the intersection point \eqref{eaten:ab}, the string ansatz for the phases can be written for any integer pair $(m_1',m_2')$ as
	\begin{equation}
		a=f_a(n_1m_1'+n_2m_2')\theta,\quad b=- f_a\frac{n_2m_1'f_1^2-n_1m_2'f_2^2}{f_1f_2}\theta,\label{eaten:ansatz1}
	\end{equation}
	where $\theta$ is the angular coordinate in the plane orthogonal to the string. We refer to the strings associated with the Bézout coefficients $(m_1',m_2')$ as fundamental axion strings. These fundamental strings provide explicit examples of sections of the axion orbit, as discussed at the end of section~\ref{sec:aorbit}.
	
	Substituting the above ansatz into Eq.~\eqref{eaten:thetaa} gives
	\begin{align}
		\theta_1=m_1'\theta,\quad 
		\theta_2=m_2'\theta.
	\end{align}
	The ansatz \eqref{eaten:ansatz1} clearly corresponds to a standard $(m_1',m_2')$ string. However, the ansatz \eqref{eaten:ansatz1} alone is not the lowest-energy configuration: far from the string core, we can eliminate the kinetic energy of $b$ by introducing a nontrivial gauge field $A_\mu$ with the following ansatz in the temporal gauge \cite{Weinberg2012}
	\begin{equation}
		A_0=0,\quad A_i=\frac{f_a}{\mu_A}\frac{n_2m_1'f_1^2-n_1m_2'f_2^2}{f_1f_2}h(r)\partial_i\theta,
	\end{equation}
	where $r$ is the distance to the string core and the coefficient function $h(r)$ satisfies
	\begin{equation}
		h(0)=0,\quad h(\infty)=1.
	\end{equation}
	Thus, the axion string is a hybrid string with a global component $a$ and a local component consisting of $b$ and $A_\mu$.
	Because of the nontrivial gauge field configuration, the hybrid axion string carries the gauge flux
	\begin{equation}
		\oint\mathrm{d}x_i \; A_i=2\pi\frac{f_a}{\mu_A}\frac{n_2m_1'f_1^2-n_1m_2'f_2^2}{f_1f_2}.
	\end{equation}
	Because $n_1$ and $n_2$ are nonzero, vanishing gauge flux for a nontrivial string requires both $m_1'$ and $m_2'$ to be nonzero and
	\begin{equation}
		\frac{n_1m_2'}{n_2m_1'}=\frac{f_1^2}{f_2^2}.\label{eaten:zeroflux}
	\end{equation}
	For fundamental axion strings, this condition can not be satisfied, so the gauge flux is necessarily nonzero. For non-fundamental axion strings, vanishing gauge flux is also difficult to realize, since the left-hand side (LHS) of Eq.~\eqref{eaten:zeroflux} is rational, whereas the RHS is generically irrational.
	
	After pointwise gauge identification, the contribution from $b$ is eliminated, leaving only the following field configuration on the quotient orbit:
	\begin{equation}
			a=f_a(n_1m_1'+n_2m_2')\theta,
	\end{equation}
	with period $2\pi F_a$. This is similar to choosing the unitary gauge. However, if we explicitly impose the unitary gauge, i.e.,
	\begin{equation}
		b\rightarrow b+f_a\frac{n_2m_1'f_1^2-n_1m_2'f_2^2}{f_1f_2}\theta,\quad A_i \rightarrow A_i-\frac{f_a}{\mu_A}\frac{n_2m_1'f_1^2-n_1m_2'f_2^2}{f_1f_2}\partial_i\theta,
	\end{equation}
	it can be verified that this gauge transformation is singular at $r=0$ because it changes the winding number. This indicates that the contribution of $b$ is not entirely a gauge redundancy. When we eliminate $b$ through pointwise gauge identification, we also eliminate physical information such as gauge flux. 
	
	Therefore, we cannot reduce the hybrid axion string to a ``pure'' axion string with no contribution from $b$. The same reasoning explains why the ``global'' axion strings in the PQWW model \cite{Dvali1993} or DFSZ-type models \cite{Dvali2025,Choi2026} generically carry gauge flux: these so-called ``global'' strings are actually hybrid strings with contributions from gauge fields. The presence of gauge flux constitutes one of the fundamental differences between this two-phase model and a single-phase model at the level of topological defects.
	
	When the potential \eqref{eaten:Va} turns on, a hybrid axion string with winding number $(m_1',m_2')$ is attached to DWs with DW number
	\begin{equation}
		N_{DW}^{(m_1',m_2')}=|n_1m_1'+n_2m_2'|.
	\end{equation}
	In particular, for the fundamental axion strings, the corresponding DW number is
	\begin{equation}
		N_{DW}=\operatorname{gcd}(n_1,n_2),
	\end{equation}
	which agrees with the value of $N_{DW}$ in \eqref{eaten:NDW}.

	\section{Mixing with a heavier phase field}
	\label{sec:heavy}
	In this section, we also consider a simplified two-phase model. The two phase fields $\phi_1,\phi_2$ and the corresponding dimensionless fields $\theta_1,\theta_2$ are defined as in Eqs.~\eqref{eaten:phi12} and \eqref{eaten:thetaphi}. The Lagrangian is
	\begin{equation}
		\mathcal{L}=\sum_{i=1}^2\frac{f_i^2}{2}\partial_\mu\theta_i\partial^\mu\theta_i-V(\theta_1,\theta_2),\label{heavy:Lorigin}
	\end{equation}
	where the scalar potential takes the form
	\begin{equation}
		V(\theta_1,\theta_2)=\Lambda_1^4[1-\cos(n_1\theta_1+n_2\theta_2)]+\Lambda_2^4[1-\cos(m_1\theta_1+m_2\theta_2)],\label{heavy:Vtheta}
	\end{equation}
	which is the special case of Eq.~\eqref{intro:Vtheta} with $N=n=2$. Here, the coefficients $n_1,n_2$ and $m_1,m_2$ are all integers. 
	\subsection{The vacua and the mass eigenstates}
	\subsubsection{The number of vacua}
	We first consider the number of vacua in the potential \eqref{heavy:Vtheta}. If $(\theta_1,\theta_2)$ is a vacuum, there exists an integer pair $(n,m)$ satisfying
	\begin{equation}
		\begin{pmatrix}
			n_1&n_2\\
			m_1&m_2
		\end{pmatrix}\begin{pmatrix}
			\theta_1\\\theta_2
		\end{pmatrix}=2\pi\begin{pmatrix}
			n\\
			m
		\end{pmatrix}.
	\end{equation}
	Thus, each integer pair $(n,m)$ corresponds to a vacuum.
	We denote the matrix by
	\begin{equation}
		A\equiv\begin{pmatrix}
			n_1&n_2\\
			m_1&m_2
		\end{pmatrix},\label{heavy:A}
	\end{equation}
	with determinant
	\begin{equation}
		D\equiv n_1m_2-n_2m_1.
	\end{equation}
	If $D$ is zero, one of the mass eigenstates is massless. Therefore, we require $D\neq 0$. Two different integer pairs $(n,m)$ and $(n',m')$ can correspond to the same vacuum, provided that there exists another integer pair $(p,q)$ satisfying
	\begin{equation}
		\begin{pmatrix}
			n\\
			m
		\end{pmatrix}-\begin{pmatrix}
		n'\\
		m'
		\end{pmatrix}=A\begin{pmatrix}
			p\\q
		\end{pmatrix}.
	\end{equation} 
	This condition defines an equivalence relation on $\mathbb{Z}^2$. The number of vacua is the number of inequivalent integer pairs, which is the cardinality of the quotient set
	\begin{equation}
		N_{\text{UV}}=|\mathbb{Z}^2/(A\mathbb{Z}^2)|=|D|.
	\end{equation}
	
	\subsubsection{The mass eigenstates}
	The mass matrix obtained from the potential \eqref{heavy:Vtheta} is 
	\begin{equation}
		\mathcal{M}^2=\begin{pmatrix}
			\alpha&\beta\\
			\beta&\eta
		\end{pmatrix}\equiv\begin{pmatrix}
			\frac{\Lambda_1^4 n_1^2+\Lambda_2^4 m_1^2}{f_1^2}&\frac{\Lambda_1^4 n_1 n_2+\Lambda_2^4 m_1m_2}{f_1f_2}\\
			\frac{\Lambda_1^4 n_1 n_2+\Lambda_2^4 m_1m_2}{f_1f_2}&\frac{\Lambda_1^4 n_2^2+\Lambda_2^4 m_2^2}{f_2^2}
		\end{pmatrix}.
	\end{equation}
	Because $D\neq0$, the mass eigenvalues are both positive
	\begin{align}
		m_a^2=\frac{\alpha+\eta-\sqrt{(\alpha-\eta)^2+4\beta^2}}{2},\quad m_b^2=\frac{\alpha+\eta+\sqrt{(\alpha-\eta)^2+4\beta^2}}{2}.\label{heavy:ma}
	\end{align}
	
	In this model, we require the mass hierarchy
	\begin{equation}
		m_a\ll m_b.\label{heavy:masshierarchy}
	\end{equation}
	For $\beta\neq0$, the relation between $(\theta_1,\theta_2)$ and the mass eigenstates $(a,b)$ corresponding to $m_a$ and $m_b$, respectively, is 
	\begin{equation}
		\begin{pmatrix}
			\theta_1\\\theta_2
		\end{pmatrix}=\begin{pmatrix}
			\frac{\beta}{f_1\rho_a}&\frac{\beta}{f_1\rho_b}\\
			\frac{m_a^2-\alpha}{f_2\rho_a}&\frac{m_b^2-\alpha}{f_2\rho_b}
		\end{pmatrix}\begin{pmatrix}
			a\\b
		\end{pmatrix}.\label{heavy:thetaphi}
	\end{equation}
	where
	\begin{equation}
		\rho_a=\sqrt{\beta^2+(m_a^2-\alpha)^2},\quad \rho_b=\sqrt{\beta^2+(m_b^2-\alpha)^2}.
	\end{equation}
	
	This relation implies that the slopes of the $a$-orbit and $b$-orbit are 
	\begin{equation}
		s_a=\frac{f_1}{f_2}\frac{m_a^2-\alpha}{\beta},\quad s_b=\frac{f_1}{f_2}\frac{m_b^2-\alpha}{\beta}.
	\end{equation}
	For generic values of the real parameters $s_a$ and $s_b$, the $a$-orbit and $b$-orbit are generically not closed. If we identify the lighter mass eigenstate $a$ as the axion field, both the period and $N_{DW}$ are generically ill-defined. We will show that $a$ coincides locally with the axion near the vacuum in section~\ref{sec:can}; however, generically this identification does not hold globally.

	The potential \eqref{heavy:Vtheta} in terms of the mass eigenstates is
	\begin{equation}
		V(a,b)=\Lambda_1^4\Big[1-\cos\Big(\frac{a}{f_{a,n}}+\frac{b}{f_{b,n}}\Big)\Big]+\Lambda_2^4[1-\cos\Big(\frac{a}{f_{a,m}}+\frac{b}{f_{b,m}}\Big)\Big],\label{heavy:Vab}
	\end{equation}
	where
	\begin{align}
		f_{a,n}&=\operatorname{sgn}(\beta)\frac{\sqrt{f_1^2+f_2^2s_a^2}}{n_1+n_2s_a},\quad 	f_{b,n}=\operatorname{sgn}(\beta)\frac{\sqrt{f_1^2+f_2^2s_b^2}}{n_1+n_2s_b},\label{heavy:fn}\\
		f_{a,m}&=\operatorname{sgn}(\beta)\frac{\sqrt{f_1^2+f_2^2s_a^2}}{m_1+m_2s_a},\quad 
		f_{b,m}=\operatorname{sgn}(\beta)\frac{\sqrt{f_1^2+f_2^2s_b^2}}{m_1+m_2s_b}.\label{heavy:fm}
	\end{align}
	Here, $\operatorname{sgn}(\beta)$ denotes the sign of $\beta$.
	Since $a$ and $b$ are mass eigenstates, the following identity holds
	\begin{equation}
		\frac{\Lambda_1^4}{f_{a,n}f_{b,n}}+\frac{\Lambda_2^4}{f_{a,m}f_{b,m}}=0.\label{heavy:xi0}
	\end{equation}
	Note that the numerators in Eqs.~\eqref{heavy:fn} and \eqref{heavy:fm} are always nonzero, whereas the denominators can vanish for special choices of parameters. Although such cases are difficult to realize, since they require either $s_a$ or $s_b$ to be rational, we briefly discuss them here for completeness. For example, suppose that the denominator of $f_{a,n}$ vanishes, i.e.
	\begin{equation}
		n_1+n_2s_a=0.
	\end{equation}
	Together with Eq.~\eqref{heavy:xi0} and $D\neq0$, this implies
	\begin{equation}
		m_1+m_2s_b=0.
	\end{equation}
	Thus, the denominator of $f_{b,m}$ also vanishes. Consequently, $f_{a,n}$ and $f_{b,m}$, as defined in Eqs.~\eqref{heavy:fn} and \eqref{heavy:fm}, formally diverge. However, only their inverses, $1/f_{a,n}$ and $1/f_{b,m}$, enter the potential in Eq.~\eqref{heavy:Vab}. In this special case,
	\begin{equation}
		\frac{1}{f_{a,n}}=\frac{1}{f_{b,m}}=0,
	\end{equation}
	so the mass eigenstates $a$ and $b$ are decoupled in the potential. After integrating out $b$ through its EOM with the derivative term neglected, as discussed in section~\ref{sec:oL}, the axion coincides globally with $a$.
	Accordingly, the axion orbit coincides with the $a$-orbit, which, because $s_a$ is rational, is closed and has a well-defined period and $N_{DW}$. In the remainder of this section, we exclude these special cases, so the quantities in Eqs.~\eqref{heavy:fn} and \eqref{heavy:fm} are finite and well defined.
	
	For $\beta=0$, the mass eigenstates are simply $\phi_1$ and $\phi_2$. For example, if $a=\phi_1$ and $b=\phi_2$, then
	\begin{equation}
		f_{a,n}=\frac{f_1}{n_1},\quad f_{b,n}=\frac{f_2}{n_2},\quad f_{a,m}=\frac{f_1}{m_1},\quad f_{b,m}=\frac{f_2}{m_2}.
	\end{equation}
	Therefore, the form of the potential in Eq.~\eqref{heavy:Vab} remains valid in the  $\beta=0$ case.
	
	Similar to the $\beta\neq0$ scenario, the zero denominators in the above expressions only imply that $a$ and $b$ are decoupled in the potential. These special cases are excluded in the rest of this section.
	
	\subsection{The axion orbit and effective Lagrangian}\label{sec:oL}

	\subsubsection{The expansion of the effective Lagrangian}\label{sec:expansion}
	Because of the mass hierarchy in Eq.~\eqref{heavy:masshierarchy}, one can integrate out the heavier field $b$ to obtain an effective theory at the energy scale $E\ll m_b$. At tree level, the effective Lagrangian can be obtained by substituting the solution of the EOM for $b$ into the original Lagrangian:
	\begin{equation}
		b=-\frac{1}{\square+m_b^2}\frac{\partial \tilde{V}}{\partial b},\label{heavy:EOM1}
	\end{equation}
	where $\square\equiv\partial^\mu\partial_\mu$ is the d'Alembert operator and  
	\begin{equation}
		\tilde{V}(a,b)=V(a,b)-\frac{1}{2}m_b^2b^2.
	\end{equation}
	We can expand the operator on the RHS of Eq.~\eqref{heavy:EOM1} in powers of $\square/m_b^2$
	\begin{equation}
		\frac{1}{\square+m_b^2}=\frac{1}{m_b^2}-\frac{\square}{m_b^4}+\cdots.\label{heavy:square}
	\end{equation}
	Note that $\square/m_b^2\sim E^2/m_b^2$. If the solution of the EOM can also be expanded in powers of $E^2/m_b^2$
	\begin{equation}
		b(x)=b_0(x)+b_1(x)+\cdots,\label{heavy:bb0}
	\end{equation}
	with $b_{n+1}/b_{n}\sim E^2/m_b^2$, the EOM \eqref{heavy:EOM1} can be solved order by order in $E^2/m_b^2$. In particular, the zeroth-order and first-order EOMs are
	\begin{equation}
		\frac{\partial V}{\partial b}\Big|_{b=b_0}=0,\quad b_1=-\Big(\frac{\partial^2 V}{\partial b^2}\Big|_{b=b_0}\Big)^{-1}\square b_0.\label{heavy:EOM2}
	\end{equation}
	We refer to the expansions of the operator in Eq.~\eqref{heavy:square} and of 
	$b$ in Eq.~\eqref{heavy:bb0} collectively as the derivative expansion.
	Then, to the order $E^2/m_b^2$, the tree-level effective Lagrangian is
	\begin{equation}
		\mathcal{L}_{\text{eff}}=\frac{1}{2}\partial_\mu a\partial^\mu a+\frac{1}{2}\partial_\mu b_0\partial^\mu b_0-V(a,b_0).\label{heavy:Leff}
	\end{equation}
	In section~\ref{sec:validity}, we will discuss the validity of this effective Lagrangian and the conditions under which the expansions break down.

	As the solution of the zeroth-order EOM in Eq.~\eqref{heavy:EOM2}, $b_0$ depends on $a$, i.e., $b_0=b_0(a)$.
	The Lagrangian \eqref{heavy:Leff} is the restriction of the original Lagrangian \eqref{heavy:Lorigin} from the torus to the trajectories $\big(a,b_0(a)\big)$. These trajectories form the field space of the effective Lagrangian. 
	
	More formally, we can define
	\begin{equation}
		F(a,b)\equiv\frac{\partial V}{\partial b},
	\end{equation}
	then
	\begin{equation}
		F(a,b)=\frac{\Lambda_1^4}{f_{b,n}}\sin\Big(\frac{a}{f_{a,n}}+\frac{b}{f_{b,n}}\Big)+\frac{\Lambda_2^4}{f_{b,m}}\sin\Big(\frac{a}{f_{a,m}}+\frac{b}{f_{b,m}}\Big).
	\end{equation}
	Since it contains no derivative operators, the zeroth-order EOM reduces to an algebraic constraint $F(a,b)=0$.
	The points satisfying this constraint form the preimage
	\begin{equation}
		S\equiv F^{-1}(0).
	\end{equation}
	Below, we will show that $S$ is compact, so each of its connected components is also compact. Furthermore, we can divide $S$ into two sets according to the type of connected component:
	\begin{equation}
		S=S_{\text{vac}}\bigcup S_{\text{novac}},
	\end{equation}
	where each connected component of $S_{\text{vac}}$ contains at least one vacuum, whereas $S_{\text{novac}}$ contains no vacuum. If the Lagrangian \eqref{heavy:Leff} is valid, $S_{\text{vac}}$ is the effective field space. The finiteness of $N_{UV}$ ensures that $S_{\text{vac}}$ is compact.

	Except for some special parameter values, we will show that $S_{\text{vac}}$ is a union of disjoint non-contractible loops on the torus, i.e.,
	\begin{equation}
		S_{\text{vac}}\cong\bigcup_{i} S^1_i.\label{heavy:SS1}
	\end{equation}
	We identify the effective field on one of these loops as the axion, and the corresponding loop as an axion orbit.
	
	The above discussion of the effective field space is similar to that in Ref.~\cite{Berg2009}. However, a significant difference between the two analyses concerns the compactness of the UV field space. In Ref.~\cite{Berg2009}, the authors introduce the axion monodromy potential $W(r)=\frac{1}{2}m^2r^2$, which is inconsistent with the periodicity of the axion $r$. Thus, the physical UV field space in Ref.~\cite{Berg2009} is a cylinder $\mathbb{R}\times S^1$ rather than a torus $S^1\times S^1$.
	
	\subsubsection{The parametrization of the axion orbit}\label{sec:para}
	We first change the variables of $F$ from $(a,b)$ to $(\theta_1,\theta_2)$
		\begin{equation}
		F(\theta_1,\theta_2)=\frac{\Lambda_1^4}{f_{b,n}}\sin(n_1\theta_1+n_2\theta_2)+\frac{\Lambda_2^4}{f_{b,m}}\sin(m_1\theta_1+m_2\theta_2).\label{heavy:Ftheta}
	\end{equation}
	It can be verified that $F(\theta_1,\theta_2)$ is a smooth function on the torus. By continuity and the fact that $\{0\}$ is a closed set in $\mathbb{R}$, $S$ is a closed set. Furthermore, since the torus is compact, the closed set $S$ is also compact. 
	
	Periodicity requires not only that the field space be compact but also that it be homeomorphic to $S^1$. 
	Therefore, we next show that each physical branch in $S$ is homeomorphic to $S^1$. To this end, we define a new basis 
	\begin{equation}
		\begin{pmatrix}
			X\\Y
		\end{pmatrix}=A\begin{pmatrix}
			\theta_1\\\theta_2
		\end{pmatrix},\label{heavy:defXY}
	\end{equation}
	where $A$ is defined in Eq.~\eqref{heavy:A}. Substituting this relation into Eq.~\eqref{heavy:Ftheta} gives
	\begin{equation}
		F(X,Y)=\frac{\Lambda_1^4}{f_{b,n}}\sin X+\frac{\Lambda_2^4}{f_{b,m}}\sin Y.
	\end{equation}
	Note that the vacua of the potential \eqref{heavy:Vtheta} correspond to $(X,Y)=(2\pi n,2\pi m)$. It can be verified that every vacuum satisfies the constraint $F(X,Y)=0$, so all vacua are contained in $S$.
	
	The advantage of $(X,Y)$ is that, depending on the parameters, the constraint can be solved for $X$ or $Y$.
	\begin{itemize}
		\item $|f_{a,n}|<|f_{a,m}|$\\
		In this scenario, we can define
		\begin{equation}
			\xi\equiv\frac{f_{a,n}}{f_{a,m}}=-\frac{\Lambda_1^4f_{b,m}}{\Lambda_2^4f_{b,n}},
		\end{equation}
		where the last equality follows directly from Eq.~\eqref{heavy:xi0}. In this scenario, $|\xi|<1$, so the constraint $F(X,Y)=0$ has the solutions 
		\begin{equation}
			Y=\arcsin(\xi\sin X)+2\pi m, \label{heavy:YX}
		\end{equation}
		or
		\begin{equation}
			Y=\pi-\arcsin(\xi\sin X)+2\pi m.\label{heavy:YpiX}
		\end{equation}
		For a fixed $m$, Eq.~\eqref{heavy:YX} and Eq.~\eqref{heavy:YpiX} each describe a curve on the torus, with the respective parametric equations
		\begin{equation}
			\begin{pmatrix}
				\theta_1\\\theta_2
			\end{pmatrix}=A^{-1}\begin{pmatrix}
				X\\ g(X)+2\pi m
			\end{pmatrix}\label{heavy:Xparameter}
		\end{equation}
		and
		\begin{equation}
			\begin{pmatrix}
				\theta_1\\\theta_2
			\end{pmatrix}=A^{-1}\begin{pmatrix}
			X\\ \pi-g(X)+2\pi m
		\end{pmatrix},\label{heavy:Xparameter1}
		\end{equation}
		where $g(X)$ is defined as
		\begin{equation}
			g(X)\equiv\arcsin(\xi\sin X),\quad g'(X)\equiv\frac{\mathrm{d}}{\mathrm{d}X}g(X).\label{heavy:gX}
		\end{equation}
		Since the tangent vectors of these two parametric equations are always nonzero, the parametrizations in \eqref{heavy:Xparameter} and \eqref{heavy:Xparameter1} are regular.
	
		First, we show that the branches \eqref{heavy:Xparameter} and \eqref{heavy:Xparameter1} do not intersect. Consider the first branch with integer $m$ and parameter $X$, and the second branch with integer $m'$ and parameter $X'$. They intersect if there exists an integer pair $(p,q)$ satisfying
			\begin{align}
			X-X'&=2\pi(n_1p+n_2q),\label{heavy:XminusX0}\\
			g(X)+g(X')-\pi+2\pi(m-m')&=2\pi (m_1p+m_2q).\label{heavy:gXminusgX0}
		\end{align}
		The condition \eqref{heavy:XminusX0} leads to
		\begin{equation}
			g(X)=g(X').
		\end{equation}
		Then, Eq.~\eqref{heavy:gXminusgX0} reduces to
		\begin{equation}
			2g(X)-\pi=2\pi k,\quad k=m_1p+m_2q-m+m'.\label{heavy:2gX}
		\end{equation}
		It can be verified that 
		\begin{equation}
			-2\pi<2g(X)-\pi<0.
		\end{equation}
		Thus, the condition \eqref{heavy:2gX} cannot be satisfied, which implies that the curves from Eq.~\eqref{heavy:Xparameter} cannot intersect those from Eq.~\eqref{heavy:Xparameter1}. Together with the fact that the branch \eqref{heavy:Xparameter1} contains no vacua, this implies that the only physical branch is Eq.~\eqref{heavy:Xparameter}. 
	
		Next, we prove that a single curve described by Eq.~\eqref{heavy:Xparameter} is closed and does not intersect itself. If two different parameter values $X$ and $X'$ correspond to the same point on the torus, there exists an integer pair $(p,q)$ satisfying the following conditions:
		\begin{align}
			X-X'&=2\pi(n_1p+n_2q),\label{heavy:XminusX}\\
			g(X)-g(X')&=2\pi (m_1p+m_2q).\label{heavy:gXminusgX}
		\end{align}
		Since $g(X)=g(X')$, the condition \eqref{heavy:gXminusgX} implies that the integers $p$ and $q$ can only take the following values
		\begin{equation}
			p=\frac{m_2 k}{\operatorname{gcd}(m_1,m_2)},\quad q=-\frac{m_1 k}{\operatorname{gcd}(m_1,m_2)},
		\end{equation}
		where $k$ is an arbitrary integer. Substituting this result into Eq.~\eqref{heavy:XminusX} gives
		\begin{equation}
			X-X'=\operatorname{sgn}(D)k T_X,\label{heavy:XminusX1}
		\end{equation}
		where
		\begin{equation}
			T_X\equiv 2\pi \frac{|D|}{\operatorname{gcd}(m_1,m_2)}.
		\end{equation}
		Eq.~\eqref{heavy:XminusX1} implies that the fundamental period of the parameter $X$ is $T_X$. Within a single period $[X,X+T_X)$, there is no intersection point. Therefore, for a given $m$, the curve from Eq.~\eqref{heavy:Xparameter} is homeomorphic to $S^1$. Furthermore, considering a trajectory from $X=0$ to $X=T_X$, the winding number is (omitting the overall sign $\operatorname{sgn}(D)$)
		\begin{equation}
			\Big(\frac{m_2}{\operatorname{gcd}(m_1,m_2)},\frac{- m_1}{\operatorname{gcd}(m_1,m_2)}\Big),
		\end{equation}
		which is nonzero. Thus, each closed curve is a non-contractible loop on the torus.

		Finally, we show that, for different values of $m$, the curves from Eq.~\eqref{heavy:Xparameter} are either identical or disjoint. Consider two curves with $m$ and $m'$. If they pass through the same point on the torus, there exists an integer pair $(p,q)$ satisfying
		\begin{align}
		X-X'&=2\pi(n_1p+n_2q),\label{heavy:XminusX2}\\
		g(X)-g(X')+2\pi (m-m')&=2\pi (m_1p+m_2q),\label{heavy:gXminusgX1}
		\end{align}
		where $X$ and $X'$ are the parameters for the first and second curves, respectively. Again, since $g(X)=g(X')$, Eq.~\eqref{heavy:gXminusgX1} gives
		\begin{equation}
			m-m'=m_1p+m_2q\in\operatorname{gcd}(m_1,m_2)\mathbb{Z}.
		\end{equation}
		This defines an equivalence relation on $\mathbb{Z}$. The subgroup $\operatorname{gcd}(m_1,m_2)\mathbb{Z}$ partitions $\mathbb{Z}$ into disjoint cosets. Only if $m$ and $m'$ belong to the same coset can the curves pass through the same point. Furthermore, if $m$ and $m'$ belong to the same coset, using Eqs.~\eqref{heavy:XminusX2}--\eqref{heavy:gXminusgX1}, the parametric equation for $m$ can be rewritten as
		\begin{align}
			\begin{pmatrix}
				\theta_1\\\theta_2
			\end{pmatrix}=A^{-1}\begin{pmatrix}
			X'\\ g(X')+2\pi m'
		\end{pmatrix}+2\pi\begin{pmatrix}
		p\\q
		\end{pmatrix}.
		\end{align}
		Thus, the images of the parametric equations for $m$ and $m'$ are identical, i.e., the two curves are the same. Therefore, the number of distinct closed curves
		equals the number of distinct cosets
		\begin{equation}
				N_{\text{orbit}}=|\mathbb{Z}/\operatorname{gcd}(m_1,m_2)\mathbb{Z}|=\operatorname{gcd}(m_1,m_2).
		\end{equation}	
		Taken together, these results imply that, in this scenario, $S_{\text{vac}}$ is the union of $N_{\text{orbit}}$ disjoint non-contractible loops, i.e.,
		\begin{equation}
			S_{\text{vac}}\cong\bigcup_{i=1}^{N_{\text{orbit}}} S^1_i.
		\end{equation}
		
		\item $|f_{a,m}|<|f_{a,n}|$\\
		This is the ``mirror'' scenario of $|f_{a,n}|<|f_{a,m}|$. We can always relabel $(\Lambda_1^4,\Lambda_2^4)$ as $(\Lambda_2^4,\Lambda_1^4)$, and $(n_1,n_2,m_1,m_2)$ as $(m_1,m_2,n_1,n_2)$. This relabeling simply exchanges the first and second terms of the potential \eqref{heavy:Vtheta}, so the mass matrix remains unchanged. If $|f_{a,m}|<|f_{a,n}|$ before relabeling, then $|f_{a,n}|<|f_{a,m}|$ after relabeling.
		
		Therefore, all the results in this scenario, such as the parametric equation, the period of the parameter, $N_{\text{orbit}}$ and $N_{DW}$, can be obtained from the preceding scenario by relabeling the parameters. 
	
		\item $|f_{a,n}|=|f_{a,m}|$\\
		In this degenerate scenario, the constraint $F(X,Y)=0$ has the solutions
		\begin{equation}
			Y=\xi X+2\pi m,\label{heavy:YxiX}
		\end{equation}
		or
		\begin{equation}
			Y=\pi-\xi X+2\pi m,\label{heavy:YpixiX}
		\end{equation}
		where $\xi=f_{a,n}/f_{a,m}$ is $1$ or $-1$.
		
		A significant difference from the two preceding scenarios is that each curve described by Eq.~\eqref{heavy:YxiX} intersects every curve described by Eq.~\eqref{heavy:YpixiX}. Thus, $S$ has only one connected component in this scenario.
		Furthermore, $\cos X$ vanishes at the intersection points. Consequently, $M_b^2(X)$ defined in Eq.~\eqref{heavy:MbX} also vanishes there, and the derivative expansion in Eq.~\eqref{heavy:bb0} breaks down at these points.

	 	For $\beta\neq0$, the degeneracy of $|f_{a,n}|$ and $|f_{a,m}|$ implies that \footnote{	For either alternative in Eq.~\eqref{heavy:sa=}, the corresponding denominator cannot vanish when that alternative is realized, since this would also require the corresponding numerator to vanish, implying $D=0$.}
	 	\begin{align}
	 		s_a=-\frac{n_1-m_1}{n_2-m_2}\quad \text{or}\quad s_a=-\frac{n_1+m_1}{n_2+m_2}.\label{heavy:sa=}
	 	\end{align}
	 	In both equations above, the LHS is a generic real number, whereas the RHS is rational. In the absence of additional constraints or parameter fine-tuning, the above conditions are difficult to realize.
	 	
	 	For $\beta=0$, the degeneracy implies
	 	\begin{equation}
	 		|n_1|=|m_1|\quad \text{or}\quad |n_2|=|m_2|.
	 	\end{equation}  
	 	In addition, $\beta=0$ itself implies
	 	\begin{equation}
	 		\Lambda_1^4 n_1 n_2+\Lambda_2^4 m_1m_2=0.
	 	\end{equation}
	 	As in the $\beta\neq0$ scenario, the above conditions are difficult to satisfy because $\Lambda_1^4/\Lambda_2^4$ is a generic real number.
	\end{itemize}
	Since the second, mirror scenario can be mapped to the first and the third, degenerate scenario is difficult to realize and does not admit a globally valid derivative expansion, we discuss only the first scenario in the remainder of this paper.
	\subsubsection{The effective Lagrangian}
	From the definition \eqref{heavy:defXY} of $(X,Y)$, one can obtain the relation between $(X,Y)$ and $(a,b)$
	\begin{align}
	\begin{pmatrix}
		X\\Y
	\end{pmatrix}=B\begin{pmatrix}
		a\\b
	\end{pmatrix},\label{heavy:XYab}
	\end{align}
	where
	\begin{equation}
		B=\begin{pmatrix}
			1/f_{a,n}&1/f_{b,n}\\
			1/f_{a,m}&1/f_{b,m}
		\end{pmatrix}.
	\end{equation}
	Substituting the constraint \eqref{heavy:YX} into the relation above gives
	\begin{equation}
		\begin{pmatrix}
			a\\b
		\end{pmatrix}=B^{-1}\begin{pmatrix}
		X\\g(X)+2\pi m
	\end{pmatrix}.\label{heavy:abX}
	\end{equation}
	We denote this relation by $\big(a(X),b(X)\big)$. One can verify that the zeroth-order EOM of $b$
	\begin{equation}
		F\big(a(X),b(X)\big)=0
	\end{equation}
	is automatically satisfied. Thus, the effective Lagrangian \eqref{heavy:Leff} can be parameterized by $X$
	\begin{equation}
		\mathcal{L}_{\text{eff}}=\frac{1}{2}\partial_\mu a(X)\partial^\mu a(X)+\frac{1}{2}\partial_\mu b(X)\partial^\mu b(X)-V\big(a(X),b(X)\big).
	\end{equation}
	Substituting the explicit expression \eqref{heavy:abX} into the Lagrangian above gives 
	\begin{equation}
		\mathcal{L}_{\text{eff}}=\frac{1}{2}K(X)\partial_\mu X\partial^\mu X-V_{\text{eff}}(X),\label{heavy:Leff1}
	\end{equation}
	where
	\begin{align}
		V_{\text{eff}}(X)&=\Lambda_1^4\big(1-\cos 	X\big)+\Lambda_2^4\Big(1-\sqrt{1-\xi^2\sin^2 X}\Big),\label{heavy:Veff}\\
		K(X)&=\frac{1}{(\det B)^2}\Big[\Big(\frac{1}{f_{a,m}}-\frac{g'(X)}{f_{a,n}}\Big)^2+\Big(\frac{1}{f_{b,m}}-\frac{g'(X)}{f_{b,n}}\Big)^2\Big].\label{heavy:KX}
	\end{align}
	Note that $\mathcal{L}_{\text{eff}}$ is a periodic function: $\mathcal{L}_{\text{eff}}(X)= \mathcal{L}_{\text{eff}}(X+2\pi)$. Furthermore, it does not depend on $m$, which implies that the effective Lagrangians are identical on different axion orbits. In each $2\pi$ period, the effective potential has one true vacuum. Therefore, the number of vacua of the effective potential, or equivalently the number of DWs around an axion orbit, is
	\begin{equation}
		N_{DW}=\frac{T_X}{2\pi}=\frac{|D|}{\operatorname{gcd}(m_1,m_2)}.\label{heavy:NDW}
	\end{equation} 
	One can verify that $N_{DW}\times N_{\text{orbit}}=N_{UV}$, which is consistent with the statement at the beginning of section~\ref{sec:para}: all the vacua are included in the axion orbits.
	\subsection{The canonical axion field}
	
	\subsubsection{The length of the axion orbit}
	Once we have obtained the effective Lagrangian, we can immediately determine the length of the axion orbit which is also the fundamental period of the  canonical field defined in Eq.~\eqref{heavy:phieff3}.
	
	To determine the length of the curve, we first need to define a metric on the torus. This metric must be consistent with lengths measured in terms of the canonical fields \cite{Berg2009}
	\begin{equation}
		\mathrm{d}s^2=(\mathrm{d}\phi_1)^2+(\mathrm{d}\phi_2)^2=(\mathrm{d}a)^2+(\mathrm{d}b)^2.
	\end{equation}	
	Substituting Eq.~\eqref{heavy:abX} into the metric gives
	\begin{align}
		\mathrm{d}s^2=K(X)(\mathrm{d}X)^2,
	\end{align}
	which is consistent with the kinetic terms in the effective Lagrangian. The length of the axion orbit, denoted as $2\pi F_a$, is
	\begin{align}
	 2\pi F_a=\int_0^{T_X}\mathrm{d}X \; \sqrt{K(X)}=N_{DW}\int_0^{2\pi}\mathrm{d}X \; \sqrt{K(X)}.\label{heavy:l}
	\end{align}
	By analogy with the single-phase model, the effective axion decay constant is defined as the length between two adjacent vacua divided by $2\pi$
	\begin{equation}
		f_{a}\equiv \frac{1}{2\pi}\int_0^{2\pi}\mathrm{d}X \; \sqrt{K(X)}.\label{heavy:fa}
	\end{equation}
	Note that $K(X)$ can also be written as follows by substituting Eq.~\eqref{heavy:Xparameter} into the kinetic term:
	\begin{equation}
		K(X)=\frac{f_1^2[m_2-n_2g'(X)]^2+f_2^2[m_1-n_1g'(X)]^2}{D^2}
	\end{equation}
	Therefore, it is possible to realize $f_{a}\gg f_1,f_2$ if the average value of $\sqrt{K(X)}$ over a single period is much larger than $f_1$ and $f_2$. One example is the KNP mechanism \cite{Choi2014}, where\footnote{Note the difference in notation between Ref.~\cite{Choi2014} and this paper. The integers $m_1$ and $m_2$ in this paper correspond to $n_1$ and $n_2$ in Ref.~\cite{Choi2014}.}
	\begin{equation}
		 m_1^2\gg D^2 \quad  \text{or}\quad  m_2^2\gg D^2,
	\end{equation}
	with $|g'(X)|\ll 1$ and $f_1\sim f_2$, so that $f_a\gg f_1,f_2$.
	
	Note that both the lengths and the effective Lagrangians are identical for different axion orbits, so the axion physics is identical on these orbits.
	\subsubsection{Two types of canonical axion field}\label{sec:can}
	Since the coefficient in the kinetic term of the effective Lagrangian \eqref{heavy:Leff1} is $K(X)/2$, the parameter $X$ is not a canonical field. 
	Here, we introduce two methods for obtaining the canonical field.
	\begin{itemize}
		\item[1.] Expand the kinetic term at the vacuum\\
		In the first method, we expand the function $K(X)$ about the vacuum $X_0=2\pi n$ as
		\begin{equation}
			K(X)=K(X_0)+ \tilde{K}(\delta X),
		\end{equation}
		where
		\begin{equation}
			\tilde{K}(\delta X)=K'(X_0)\delta X+\frac{1}{2}K''(X_0)(\delta X)^2+\cdots,
		\end{equation}
		and $\delta X\equiv X-X_0$. Then, we define the canonical axion field as
		\begin{equation}
			\phi_{\text{eff}}\equiv \sqrt{K(X_0)}\delta X,\label{heavy:phieff1}
		\end{equation}
		with the period $T_X\sqrt{K(X_0)}$.
		The effective Lagrangian in terms of $\phi_{\text{eff}}$ is
		\begin{equation}
			\mathcal{L}_{\text{eff}}=\frac{1}{2}\partial_\mu \phi_{\text{eff}}\partial^\mu \phi_{\text{eff}}+\frac{1}{2K(X_0)}\tilde{K}\Big(\frac{\phi_{\text{eff}}}{\sqrt{K(X_0)}}\Big)\partial_\mu \phi_{\text{eff}}\partial^\mu \phi_{\text{eff}}-V_{\text{eff}}\Big(\frac{\phi_{\text{eff}}}{\sqrt{K(X_0)}}\Big),
		\end{equation}
		where we have used the fact that $V_{\text{eff}}(X_0+\delta X)=V_{\text{eff}}(\delta X)$.
		We can also expand the effective potential about $X_0$
		\begin{equation}
			V_{\text{eff}}\Big(\frac{\phi_{\text{eff}}}{\sqrt{K(X_0)}}\Big)=\frac{1}{2}V''_{\text{eff}}(X_0)\Big(\frac{\phi_{\text{eff}}}{\sqrt{K(X_0)}}\Big)^2+\mathcal{O}(\phi_{\text{eff}}^3),
		\end{equation} 
		then the mass of $\phi_{\text{eff}}$ is exactly
		\begin{equation}
			m_{\text{eff}}^2=\frac{V''_{\text{eff}}(X_0)}{K(X_0)},\quad \text{where} \quad V''_{\text{eff}}(X_0)=\Big(\frac{\mathrm{d}^2}{\mathrm{d}X^2}V_{\text{eff}}\Big)\Big|_{X=X_0}.
		\end{equation}
		From the expressions \eqref{heavy:Veff} and \eqref{heavy:KX}, one can verify that
		\begin{equation}
			m^2_{\text{eff}}=\frac{\Lambda_1^4}{f_{a,n}^2}+\frac{\Lambda_2^4}{f_{a,m}^2}=m_a^2.\label{heavy:meff}
		\end{equation}
		To understand this result, we expand the relation between $a$ and $X$ in Eq.~\eqref{heavy:abX} about $X_0$
		\begin{equation}
			\delta a= \frac{1}{\det B}\Big(\frac{1}{f_{b,m}}-\frac{\xi}{f_{b,n}}\Big)\delta X+\mathcal{O}[(\delta X)^2].\label{heavy:deltaa}
		\end{equation}
		where $\delta a=a(X)-a(X_0)$. Because
		\begin{equation}
			K(X_0)=\frac{1}{(\det B)^2}\Big(\frac{1}{f_{b,m}}-\frac{\xi}{f_{b,n}}\Big)^2,
		\end{equation}
		Eq.~\eqref{heavy:deltaa} implies that, to linear order near the vacuum 
		\begin{equation}
			\delta a=\phi_{\text{eff}}\quad \text{or}\quad \delta a=-\phi_{\text{eff}}.
		\end{equation}
		This explains why the mass of $a$ and that of the effective field are identical.
		
		\item[2.] Normalize the full kinetic term\\
		In the second method, we normalize the full kinetic term by defining
		\begin{equation}
			\partial_\mu \phi_{\text{eff}}\equiv \sqrt{K(X)}\partial_\mu X,
		\end{equation}
		which has the differential form
		\begin{equation}
			\mathrm{d}\phi_{\text{eff}}=\sqrt{K(X)}\mathrm{d}X.\label{heavy:phieff2}
		\end{equation}
		Note that, near the vacuum, this relation agrees with the first method in Eq.~\eqref{heavy:phieff1}.
		
		Integrating Eq.~\eqref{heavy:phieff2} gives
		\begin{equation}
			\phi_{\text{eff}}(X)=\int_{X_0}^X\mathrm{d}\tilde{X}\sqrt{K(\tilde{X})}.\label{heavy:phieff3}
		\end{equation}
		Comparing this definition with the orbit length in Eq.~\eqref{heavy:l}, we find that the canonical field in the second definition is the signed arc length along the axion orbit from $X_0$ to $X$. Thus, the period of this canonical field is $2\pi F_a$.
	
	\end{itemize}
	The first method is well suited to evaluating scattering amplitudes, whereas the second is well suited to evaluating DW configurations. In both methods, the canonical axion field agrees with the light mass eigenstate to leading order near the vacuum.
	
	So far, we have obtained the effective Lagrangian and the corresponding canonical axion. This effective Lagrangian generically differs from the standard cosine single-phase model, even when a mass hierarchy between the light and heavy fields is imposed. As an illustration, in appendix~\ref{app:DW}, we compare the corresponding DW solutions with those of a standard cosine model matched to the same axion mass, decay constant, and DW number. When the mass hierarchy originates from the hierarchy of the explicit breaking scales, the deviations are strongly suppressed. In contrast, hierarchies generated by the spontaneous breaking scales or by integer alignment can lead to appreciable modifications of the DW tension and energy-density profile. These are quantitative consequences of the induced single-phase dynamics rather than fundamental differences between two-phase and single-phase theories.
	\subsection{The anomaly couplings}
	In this subsection, we discuss the anomaly couplings in the effective Lagrangian. As an example, we consider the coupling between axions and photons in the original Lagrangian
	\begin{equation}
		\mathcal{L}\supset(k_1\theta_1+k_2\theta_2)\frac{e^2}{32\pi^2}F\tilde{F},
	\end{equation}
	where $k_1$ and $k_2$ are integers. After substituting Eq.~\eqref{heavy:Xparameter}, the coupling becomes
	\begin{equation}
		\mathcal{L}_{\text{eff}}\supset\frac{w(X)}{D}\frac{e^2}{32\pi^2}F\tilde{F},
	\end{equation}
	where
	\begin{equation}
		w(X)\equiv (k_1m_2-k_2m_1)X+(k_2n_1-k_1n_2)\big[g(X)+2\pi m\big].
	\end{equation}
	Note that $w(X)$ is consistent with the period of $X$, i.e., the difference
	\begin{equation}
		\frac{w(X+T_X)-w(X)}{D}=\operatorname{sgn}(D)\frac{k_1m_2-k_2m_1}{\gcd(m_1,m_2)}2\pi,
	\end{equation}
	is a multiple of $2\pi$.
	
	Expanding $w(X)$ about the vacuum $X_0=2\pi n$ and substituting the effective field defined in Eq.~\eqref{heavy:phieff1}, we obtain the effective axion-photon coupling at linear order
	\begin{equation}
		\mathcal{L}_{\text{eff}}\supset\frac{1}{4}g_{\phi\gamma}\phi_{\text{eff}}F\tilde{F},
	\end{equation}
	where
	\begin{equation}
		g_{\phi\gamma}=\frac{(k_1m_2-k_2m_1)+(k_2n_1-k_1n_2)\xi}{D}\frac{e^2m_{a}}{8\pi^2\sqrt{\Lambda_1^4+\Lambda_2^4\xi^2}}.\label{heavy:gphigamma}
	\end{equation}
	The expression for $g_{\phi\gamma}$ implies that the axion-photon coupling can be enhanced by 
	\begin{equation}
		(k_1m_2-k_2m_1)+(k_2n_1-k_1n_2)\xi.
	\end{equation}
	Ref.~\cite{Agrawal2018} provides an example of this enhancement. In Ref.~\cite{Agrawal2018}, $\Lambda_1\equiv\Lambda_{QCD}\ll\Lambda_2\equiv\Lambda_{H}$. The two phase fields can be identified as $\theta_1\equiv a/F_0$, $\theta_2\equiv b/F_0$. For the integer choices $n_1=0,n_2=m_1=1,m_2=M^\alpha$ and $k_1=M^\beta,k_2=0$, $g_{\phi\gamma}$ becomes
	\begin{equation}
		g_{\phi\gamma}=(-M^{\alpha+\beta}+M^{\beta}\xi)\frac{e^2m_{a}}{8\pi^2\sqrt{\Lambda_{QCD}^4+\Lambda_H^4\xi^2}}\simeq -M^{\alpha+\beta}\frac{e^2m_{a}}{8\pi^2\Lambda_{QCD}^2},
	\end{equation}
	where, in the last step, we have neglected the small quantity $\xi\sim\Lambda_{QCD}^4/\Lambda_H^4$ (see appendix~\ref{sec:numer}). This result is consistent with Ref.~\cite{Agrawal2018}: the coupling is enhanced by $M^{\alpha+\beta}$.
	
	\subsection{The validity of the effective Lagrangian}\label{sec:validity}
	In section~\ref{sec:expansion}, we use two expansions to obtain the Lagrangian \eqref{heavy:Leff}:
	\begin{itemize}
		\item[1.] Expand the action about the classical solution of the EOM of $b$, retaining only the tree-level contribution.
		\item[2.] Expand the classical solution of the EOM of $b$ in powers of $E^2/m_b^2$, omitting quadratic and higher-order terms in the effective Lagrangian.
	\end{itemize}
	Here, we investigate the validity of the approximations used above.
	
	\subsubsection{Stability of the tree-level effective Lagrangian}
	First, we discuss the stability of the tree-level effective Lagrangian. The original action $S[a,b]$ can be expanded about $b_c$
	\begin{equation}
		S[a,b]=S[a,b_c]+\frac{1}{2}\int\mathrm{d}^4x_1\mathrm{d}^4x_2\;\delta b(x_1)G^{(2)}(x_1,x_2)\delta b(x_2)+\cdots,
	\end{equation} 
	where the linear term vanishes because $b_c$ is the solution of the EOM of $b$, and
	\begin{equation}
		\delta b=b-b_c,\quad G^{(2)}(x_1,x_2)=\frac{\delta^2 S}{\delta b(x_1)\delta b(x_2)}\Big|_{b=b_c}.
	\end{equation}
	Integrating the second term by parts gives
	\begin{align}
		\frac{1}{2}\int\mathrm{d}^4x_1\mathrm{d}^4x_2\;\delta b(x_1)G^{(2)}(x_1,x_2)\delta b(x_2)=\frac{1}{2}\int\mathrm{d}^4x \; \delta b(x)O\delta b(x),
	\end{align}
	where the operator $O$ is defined as
	\begin{equation}
		O\equiv-\square-\frac{\partial^2V}{\partial b^2}\Big|_{b=b_c}.
	\end{equation}
	To assess stability against fluctuations of $b$, we Wick-rotate this operator to Euclidean space
	\begin{equation}
		O_E=-\square_E+\frac{\partial^2V}{\partial b^2}\Big|_{b=b_c}.
	\end{equation}
	The first term of $O_E$ is positive semidefinite:
	\begin{equation}
		\frac{1}{2}\int\mathrm{d}^4x \; \delta b(x)(-\square_E)\delta b(x)=\frac{1}{2}\int\mathrm{d}^4x \; \big[\partial_i\delta b(x)\big]^2\geq0,
	\end{equation}
	where $\partial_i=(\nabla,\partial_4)$ is the derivative operator in Euclidean space. Therefore, if 
	\begin{equation}
		\frac{\partial^2V}{\partial b^2}\Big|_{b=b_c}>0,\label{heavy:stable}
	\end{equation}
	the operator $O_E$ is positive definite. Provided that the derivative expansion in Eq.~\eqref{heavy:bb0} is valid, the stability condition \eqref{heavy:stable} reduces at leading order to
	\begin{equation}
		\frac{\partial^2V}{\partial b^2}\Big|_{b=b_0}>0.
	\end{equation}
	
	\subsubsection{Validity of the derivative expansion}
	Second, we discuss the validity of the derivative expansion of the EOM solution. In section~\ref{sec:expansion}, we expand $b$ as
	\begin{equation}
		b(x)=b_0(x)+b_1(x)+\cdots,
	\end{equation}
	with $b_{n+1}/b_{n}\sim E^2/m_b^2$. However, this expansion is not always valid. The EOM for $b_1$ is
	\begin{equation}
		b_1=-\Big(\frac{\partial^2 V}{\partial b^2}\Big|_{b=b_0}\Big)^{-1}\square b_0.\label{heavy:b1}
	\end{equation}
	One can verify that, near the vacuum,
	\begin{equation}
		\frac{\partial^2 V}{\partial b^2}\Big|_{b=b_0}\sim m_b^2.
	\end{equation}
	Thus, near the vacuum, $b_1/b_0$ is $\mathcal{O}(E^2/m_b^2)$. However, this relation is not guaranteed to remain valid when the effective field is far from the vacuum. In particular, if 
	\begin{equation}
		\frac{\partial^2 V}{\partial b^2}\Big|_{b=b_0}=0
	\end{equation}
	at some points on the effective axion orbit, Eq.~\eqref{heavy:b1} is singular at those points.
	
	\subsubsection{The validity criterion}
	Using Eq.~\eqref{heavy:YX}, we can parameterize
	\begin{equation}
		M^2_b(X)\equiv\frac{\partial^2 V}{\partial b^2}\Big|_{b=b_0}=\frac{\Lambda_1^4}{f^2_{b,n}}\cos X+\frac{\Lambda_2^4}{f^2_{b,m}}\sqrt{1-\xi^2\sin^2 X}.\label{heavy:MbX}
	\end{equation}
	Because the stability of the tree-level effective Lagrangian and the validity of the derivative expansion both depend on $M^2_b(X)$, we summarize the validity criterion as 
	\begin{equation}
		R_{M}\equiv\frac{M^2_{b,\text{min}}}{M^2_{b,\text{max}}}> c,
	\end{equation}
	where $0\leq c<1$ is a c-number. If $c=0$, the criterion reduces to the non-singular condition. Note that the maximum of $M^2_b(X)$ is always
	\begin{equation}
		M^2_{b,\text{max}}=m_b^2=\frac{\Lambda_1^4}{f^2_{b,n}}+\frac{\Lambda_2^4}{f^2_{b,m}}.
	\end{equation}
	However, the minimum of $M^2_b(X)$ depends on the parameters. For simplicity, we define 
	\begin{equation}
		\zeta\equiv \xi^2\frac{\Lambda_2^4}{\Lambda_1^4}.
	\end{equation}
	The expression for the minimum of $M^2_b(X)$ has two branches:
	\begin{itemize}
		\item[1.] $\Lambda_1^4\leq \Lambda_2^4$\\
		In this scenario, 
		\begin{equation}
			M^2_{b,\text{min}}=-\frac{\Lambda_1^4}{f^2_{b,n}}+\frac{\Lambda_2^4}{f^2_{b,m}}
		\end{equation}
		Thus, the criterion becomes 
		\begin{equation}
			\zeta<\frac{1-c}{1+c}.
		\end{equation}
		\item[2.] $\Lambda_2^4<\Lambda_1^4$\\
		In this scenario, 
		\begin{equation}
			M^2_{b,\text{min}}=\frac{\Lambda_2^4}{f_{b,m}^2}\sqrt{(1-\xi^2)\Big(1-\frac{\zeta^2}{\xi^2}\Big)}.
		\end{equation}
		The ratio is
		\begin{equation}
			R_{M}=\frac{1}{1+\zeta}\sqrt{(1-\xi^2)\Big(1-\frac{\zeta^2}{\xi^2}\Big)}.
		\end{equation}
	\end{itemize}
	Since the criterion for $\Lambda_1^4\leq \Lambda_2^4$ is much simpler than that for the other scenario, we consider only the first scenario in the numerical calculations in appendix~\ref{sec:numer}.
	
	We adopt $c=0.1$ as a practical, conservative choice, which ensures that $M_b^2(X)$ never falls below one tenth of its maximum along the whole orbit. With this choice and $\Lambda_1^4\leq \Lambda_2^4$, the criterion reduces to
	\begin{equation}
		\zeta<\frac{9}{11}.\label{heavy:zeta}
	\end{equation}
	
	\subsubsection{Discussion of the derivative term}
	Finally, we briefly discuss the possibility that the derivative expansion breaks down while the tree-level Lagrangian remains stable. If this expansion fails, we must consider the full EOM:
	\begin{equation}
		\square b+\frac{\partial V}{\partial b}=0.
	\end{equation}
	This equation gives the relation between the classical solution $b_c$ and $a$. If the derivative term can be omitted, this relation is algebraic, e.g., $b_c=f(a)$, and does not depend on how $a$ varies with the spacetime coordinates. However, when the derivative term cannot be omitted, this relation is functional, e.g.,
	\begin{equation}
		b_c(x)=\int\mathrm{d}y \; G(x,y) J[a(y)].
	\end{equation}  
	Only after specifying $a(x)$ can we calculate $b_c(x)$. Thus, we can only define the axion orbit as the image of the map $x \mapsto \big(a(x),b_c(x)\big)$. However, this definition faces two difficulties: (i) it is difficult to determine whether the orbit is compact, and it need not be one-dimensional; and (ii) different choices of $a(x)$ generically yield different orbits.
	
	This implies that when the derivative term cannot be neglected, the axion can no longer be described by a configuration-independent algebraic orbit. In particular, this situation may arise for nontrivial topological defects, such as DW configurations containing a non-negligible heavy-field component. In such cases, the breakdown of the single-phase reduction gives rise to a fundamental difference between the two-phase and single-phase descriptions at the level of topological defects.
	\section{The identification of the axion}\label{sec:compact}
	In this section, we first generalize the reduction mechanisms in sections~\ref{sec:eaten} and \ref{sec:heavy} from two-phase models to multi-phase models with a concise proof. We then combine these two reduction mechanisms to obtain a general route to axion identification.
	\subsection{Reduction through gauge identification}\label{sec:redu_ga}
	Consider an original Lagrangian with scalar fields $\Phi=(\phi_1,\dots,\phi_N)^T$ taking values in a compact field space $M$, such as $T^N$ or $S^N$. Suppose that there is a continuous action of the gauge group $G$ 
	\begin{equation}
		G\times M\rightarrow M,\quad (g,\Phi)\mapsto g\cdot \Phi.
	\end{equation}
	We then define the relation $\Phi\sim \Phi'$ if there exists $g\in G$ such that $\Phi'= g\cdot \Phi$. This relation can be shown to be an equivalence relation, referred to as the gauge equivalence relation. Fields that are gauge equivalent to one another form an equivalence class called a gauge orbit, denoted by $[\Phi]$. Thus, the field space after gauge identification is the quotient set $M/\sim$. With the quotient map
	\begin{equation}
		\pi: M\rightarrow M/\sim,\quad \Phi\mapsto [\Phi],
	\end{equation}
	one can define the quotient topology on $M/\sim$. The quotient field space $M/\sim$ is then compact because $M$ is compact.
	
	Moreover, in multi-phase models, the gauge group action is generally the translation of the phases with integer-valued charges. The quotient field space is then
	\begin{equation}
		T^N/U(1)^k\cong T^{N-r}.
	\end{equation}
	where $r\leq k$ is the number of $U(1)$ symmetries acting independently on the phase fields.
	If $r=N-1$, the quotient field space is homeomorphic to $S^1$. The axion is then identified as the field in this quotient field space that can locally coincide with the massive eigenstate. Its period and $N_{DW}$ can be determined from the original Lagrangian.
	\subsection{Reduction through integrating out heavy fields}\label{sec:redu_int}
	Consider an original UV Lagrangian with scalar fields $\Phi=(\phi_1,\dots,\phi_n)^T$ taking values in a compact field space, such as $T^n$ or $S^n$. We require that the Lagrangian be well-defined on the field space; thus, the potential is a smooth function.
	
	As in the model in section~\ref{sec:heavy}, we require that all the scalar fields are massive at every true vacuum, which implies that the corresponding mass matrix $\mathcal{M}^2$ is positive definite. The UV potential $V(\Phi)$ can be expanded about any vacuum $\Phi_0$ as
	\begin{equation}
		V(\Phi)=V(\Phi_0)+\frac{1}{2}(\delta\Phi)^T\mathcal{M}^2\delta\Phi+\mathcal{O}[(\delta\Phi)^3],
	\end{equation}
	where $\delta\Phi=\Phi-\Phi_0$. If $\mathcal{M}^2$ is positive definite, this expansion implies that there exists a domain containing $\Phi_0$ that contains no other vacua. This directly indicates that the UV vacua are isolated from one another. Since all true vacua have the same potential energy, we can shift this common value to zero and express the collection of vacua as
	\begin{equation}
		\mathcal{V}=V^{-1}(0),
	\end{equation}
	which is also a compact set. Since (i) all the vacua in $\mathcal{V}$ are isolated from one another and (ii) $\mathcal{V}$ is a compact set, it can be verified that the number of vacua 
	\begin{equation}
		N_{UV}=|\mathcal{V}|
	\end{equation}
	is finite.
	
	We assume that there are $m$ heavy fields with masses much larger than those of the other light massive fields.
	If the derivative terms in the EOMs of the heavy fields can be omitted, we obtain multiple algebraic constraints
	\begin{align}
		F_1(\phi_1,\dots,\phi_n)=\cdots=F_m(\phi_1,\dots,\phi_n)=0,
	\end{align}
	After integrating out the heavy fields using their EOMs, the $n-m$ effective light fields take values in the IR field space determined by the constraints above. Since $F_1,\dots,F_m$ are continuous functions, their preimages
	\begin{equation}
		S_1\equiv F_1^{-1}(0),\quad \dots,\quad  S_m\equiv F_m^{-1}(0),
	\end{equation}
	are all closed sets. The intersection of these closed sets
	\begin{equation}
		S=\bigcap_{i=1}^mS_i.
	\end{equation}
	is also closed and therefore compact. The IR field space $S_{\text{vac}}$, defined as the union of connected components of $S$ that contain at least one vacuum, is also compact because $N_{UV}$ is finite.

	In particular, as a direct generalization of the two-phase model to multi-phase models, we assume that the number of heavy fields is $n-1$ and that the potential is identical to that in Eq.~\eqref{intro:Vtheta} with $N=n$. The constraints can then be expressed as
	\begin{align}
		F_\alpha(X)=\sum_{i=1}^n A_{\alpha i}\sin X_i,\quad \alpha=1,\dots,n-1,\label{compact:FX}
	\end{align} 
	The constraints in Eq.~\eqref{compact:FX} can be regarded as linear equations in $\sin X_i$ with an $(n-1)\times n$ matrix $A$ whose entries are $A_{\alpha i}$.
	
	If all these constraints are independent, the rank of $A$ is $n-1$, which implies that the solution of Eq.~\eqref{compact:FX} can be written as 
	\begin{equation}
		(\sin X_1,\dots,\sin X_n)=t(v_1,\dots,v_n),
	\end{equation}
	where $t$ is a real parameter. 
	As discussed in section~\ref{sec:para}, we require that there exist $v_k$ satisfying
	\begin{equation}
		|v_k|>|v_i|,\quad \text{for all}\; i\neq k,
	\end{equation}
	which is the non-degeneracy condition for multi-phase models.
	Then, we can solve $X_i$ ($i\neq k$) as
	\begin{equation}
		X_i=\arcsin\Big(\frac{v_i}{v_k}\sin X_k\Big) +2\pi k_i \label{compact:Xi1}
	\end{equation}
	or
	\begin{equation}
		X_i=\pi-\arcsin\Big(\frac{v_i}{v_k}\sin X_k\Big) +2\pi k_i.\label{compact:Xi2}
	\end{equation}
	These are exactly the same formulas as those in Eqs.~\eqref{heavy:YX}-\eqref{heavy:YpiX}. 
	Thus, only the branches described by Eq.~\eqref{compact:Xi1} contribute to $S_{\text{vac}}$, which is a finite union of disjoint, non-contractible loops on $T^n$:
	\begin{equation}
		S_{\text{vac}}=\bigcup_i S^1_i.
	\end{equation}
	The axion is identified as the effective field on one of the $S^1$ orbits. This field coincides with the lightest mass eigenstate near the vacuum and has a well-defined period and $N_{DW}$.
	
	As a consequence, within the framework considered here, the compact low-energy description established for an exactly massless axion in Ref.~\cite{Fraser2020} extends to the massive axion.
	\subsection{The combination of the two reductions}	
	In the preceding discussion, we have shown that the axion can be identified in multi-phase models after gauge identification or after integrating out the heavy fields. Here, we combine these two scenarios to obtain a more general identification.
	
	We consider an $N$-phase model with $k$ independent massless phase modes that are eaten by gauge fields. Among the remaining massive modes, one is much lighter than the others.
	Under the requirements discussed in sections~\ref{sec:redu_ga} and \ref{sec:redu_int}, we can identify the axion field in two steps: (i) first perform gauge identification, reducing the model to an $(N-k)$-phase model with the quotient field space $T^{N-k}$; (ii) then integrate out the heavy fields, reducing it to a single-phase model whose effective field space is a finite union of $S^1$. We identify the axion as the effective field on one of these $S^1$ components. The corresponding $N_{DW}$, $F_a$, and the effective Lagrangian can all be calculated from the original UV Lagrangian. The axion in the standard DFSZ model is one of the best-known examples of such identification.
	
	This identification also faces the validity problem discussed in section~\ref{sec:validity}. In particular, this identification may become incomplete in nontrivial topological defects. This incompleteness leads to fundamental differences between multi-phase and single-phase models at topological defects.
	
	\section{Conclusion}\label{sec:conc}
	Confusion about the identification of the axion in multi-phase models appears in the literature. In particular, if the axion is identified as the lightest mass eigenstate, the well-definedness of the corresponding period and $N_{DW}$ can become problematic. This paper considers a class of models satisfying two conditions: (i) the massless phases are eaten by gauge fields, and (ii) there is a mass gap between the single lightest phase and the other heavy phases. 
	For a cosine-type potential and within the validity of the derivative expansion, our analysis shows that the axion can be identified as the effective field on one of the $S^1$ orbits with a well-defined period and $N_{DW}$.
	However, in the presence of topological defects, pointwise gauge identification may omit physical information, such as the gauge flux carried by axion strings, while integrating out the heavy phase fields may fail for certain DW configurations. These effects reveal fundamental differences between the multi-phase models and their reduced single-phase descriptions at the level of topological defects.

	\appendix
	
	\section{DWs in the induced and standard single-phase models}\label{app:DW}
	In this appendix, we study the axion DW configuration of the effective Lagrangian \eqref{heavy:Leff1} and compare it with that of the standard single-phase model.

	\subsection{The DW in the effective Lagrangian}\label{sec:DWeff}
	The reduction performed in section~\ref{sec:oL} gives the single-phase effective Lagrangian \eqref{heavy:Leff1}. We can then obtain the DW configuration of this Lagrangian.
	
	Here, we consider only stable DWs that are independent of time and connect two adjacent vacua. For simplicity, we also require that the DWs depend only on the coordinate $z$. The corresponding EOM is
	\begin{equation}
		K X''+\frac{1}{2}K'(X')^2=V'_{\text{eff}},\label{heavy:DWEOM2}
	\end{equation}
	where 
	\begin{equation}
		X'=\frac{\mathrm{d}}{\mathrm{d}z}X,\quad X''=\frac{\mathrm{d}^2}{\mathrm{d}z^2}X,\quad K'=\frac{\mathrm{d}}{\mathrm{d}X}K,\quad V'_{\text{eff}}=\frac{\mathrm{d}}{\mathrm{d}X}V_{\text{eff}},
	\end{equation}
	and the dependence of $K,K'$ and $V'_{\text{eff}}$ on $X$ is omitted. Using the first integral of Eq.~\eqref{heavy:DWEOM2} and the DW boundary condition, we obtain a first-order differential equation
	\begin{equation}
		\frac{1}{2}K(X')^2-V_{\text{eff}}=0.
	\end{equation}
	This gives the DW configuration in terms of $X$ in integral form 
	\begin{equation}
		z-z_0=\pm \int_{X_0}^X\mathrm{d}\tilde{X}\sqrt{\frac{K(\tilde{X})}{2V_{\text{eff}}(\tilde{X})}},\label{heavy:DWeff}
	\end{equation}
	where $X_0=X(z_0)$ and $z_0$ is the position of the DW center. The tension of this DW is
	\begin{equation}
		\sigma_{\text{eff}}=\int_{-\infty}^{+\infty}\mathrm{d}z \; \rho_{\text{eff}}=\Big|\int_{X_-}^{X_+}\mathrm{d}X\sqrt{2KV_{\text{eff}}}\Big|,
	\end{equation}
	where $X_-=X(-\infty), X_+=X(+\infty)$, and the DW energy density is $\rho_{\text{eff}}=2V_{\text{eff}}$.
	Since the integrals for the DW configuration and tension are difficult to evaluate analytically, we evaluate them numerically and present the results in section~\ref{sec:numer}.
	\subsection{The DW in the standard single-phase model}
	We can compare the DW configuration obtained in section~\ref{sec:DWeff} with that of the standard single-phase model. The standard cosine single-phase model contains three parameters: the axion mass $m_a^{\text{std}}$, the axion decay constant $f_a^{\text{std}}$, and the DW number $N_{DW}^{\text{std}}$. 
	
	For a meaningful comparison, we match these three parameters to the corresponding parameters in the effective Lagrangian (given in Eqs.~\eqref{heavy:meff}, \eqref{heavy:fa}, and \eqref{heavy:NDW}, respectively):
	\begin{equation}
		m_a^{\text{std}}=m_{a},\quad f_a^{\text{std}}=f_{a},\quad N_{DW}^{\text{std}}=N_{DW}.
	\end{equation}
	Therefore, the Lagrangian of the standard single-phase model is
	\begin{equation}
		\mathcal{L}_{\text{std}}=\frac{1}{2}\partial_\mu a\partial^\mu a-V(a),\quad V(a)=m_{a}^2f_{a}^2\Big(1-\cos\frac{a}{f_{a}}\Big).
	\end{equation}
	$N_{DW}$ does not appear explicitly in the above Lagrangian; however, it determines the period of the axion and the number of DWs attached to the axion string.
	
	The DW configuration in the standard single-phase model is
	\begin{equation}
		z-z_0=\pm\int_{a_0}^{a}\frac{\mathrm{d} \tilde{a}}{\sqrt{2V(\tilde{a})}},\label{heavy:DWa}
	\end{equation}
	where $a_0=a(z_0)$ and $z_0$ is the position of the DW center. The corresponding DW tension is $\sigma_{\text{std}}=8m_{a}f_{a}^2$.
	
	Specifically, we compare the following quantities:
	\begin{itemize}
		\item[1.] The DW tension\\
		We evaluate the deviation from unity of the ratio $R_{\sigma}$, defined as
		\begin{equation}
			R_{\sigma}\equiv \frac{	\sigma_{\text{eff}}}{\sigma_{\text{std}}}.
		\end{equation}
		\item[2.] The energy density distribution\\
		We compare the energy density distributions of DWs along the $z-$axis. These distributions are represented by $(z(X),\rho_{\text{eff}}(X))$ and $(z(a),\rho(a))$, where $z(X)$ and $z(a)$ are calculated from Eq.~\eqref{heavy:DWeff} and Eq.~\eqref{heavy:DWa}, and $\rho_{\text{eff}}(X)=2V_{\text{eff}}(X)$, $\rho(a)=2V(a)$. In particular, we can compare the ratio of the maxima of the energy density distributions
		\begin{equation}
			R_{\rho}\equiv\frac{\operatorname{Max}(\rho_{\text{eff}})}{\operatorname{Max}(\rho_{\text{std}})}.
		\end{equation}
		\item[3.] The DW configuration\\
		The DW configurations are represented by $(z(X),\phi_{\text{eff}}(X))$ and $(z(a),a)$, where $\phi_{\text{eff}}(X)$ can be defined by Eq.~\eqref{heavy:phieff1} or Eq.~\eqref{heavy:phieff3}.
		\item[4.] The potential\\
		We also compare the scalar potentials as functions of the canonical fields, represented by $(\phi_{\text{eff}}(X),V_{\text{eff}}(X))$ and $(a,V(a))$.
	\end{itemize}

	\subsection{Numerical comparisons in the three simplest realizations}\label{sec:numer}
	The mass hierarchy $m_a\ll m_b$ is required in the two-phase model. We compare the DW quantities in each of the three simplest realizations of this hierarchy. In this section, we select the branch $\Lambda_1^4\leq\Lambda_2^4$ and adopt the validity criterion in Eq.~\eqref{heavy:zeta}. In each realization, we assume that only the specified hierarchy is present, while all other independent dimensionless input ratios remain of order unity.
	
\begin{figure}[t]
		\centering
		\begin{minipage}[t]{0.3\textwidth}
			\includegraphics[width=\linewidth]{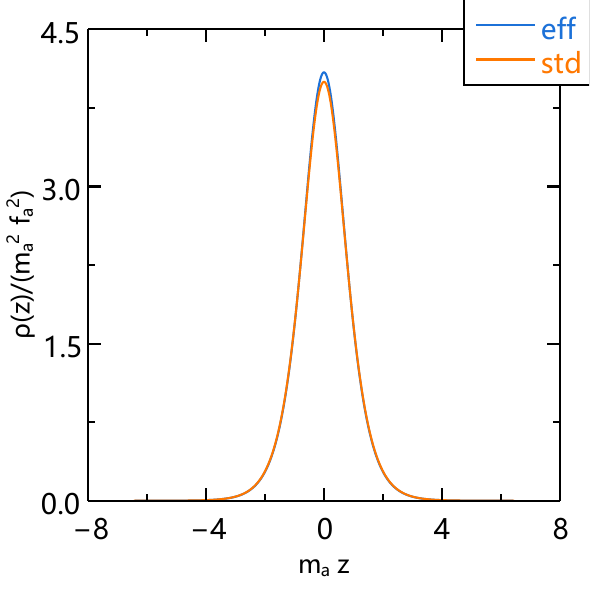}
		\end{minipage}
		\hfill
		\begin{minipage}[t]{0.3\textwidth}
			\includegraphics[width=\linewidth]{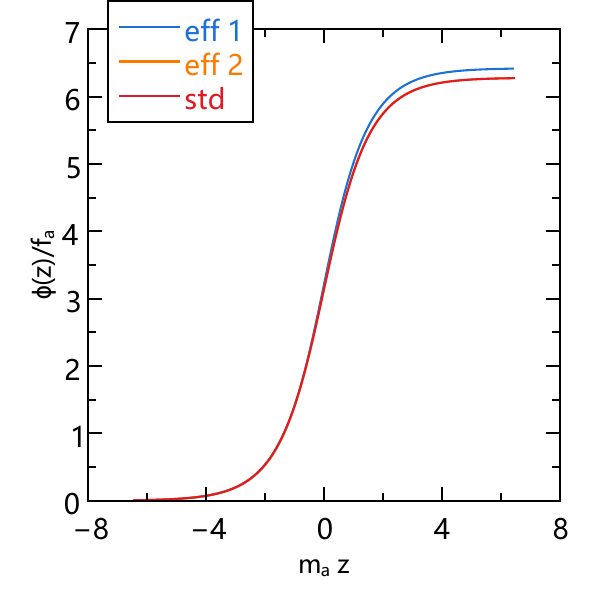}
		\end{minipage}
		\hfill
		\begin{minipage}[t]{0.3\textwidth}
			\includegraphics[width=\linewidth]{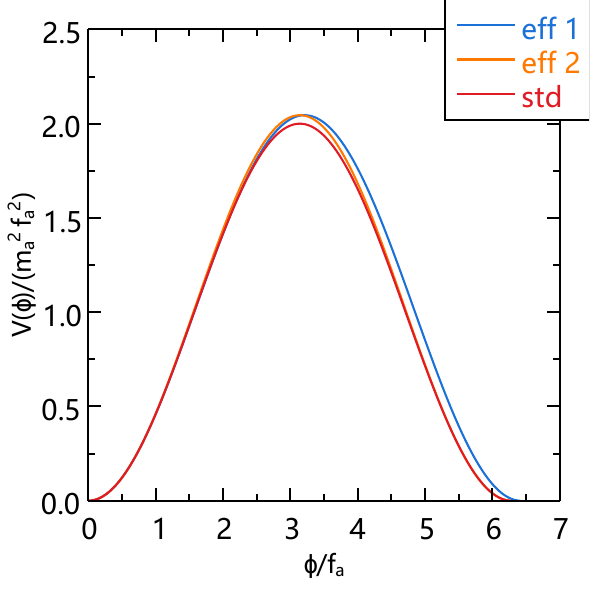}
		\end{minipage}
		
		\vspace{0.2cm} 
		
		\begin{minipage}[t]{0.3\textwidth}
			\includegraphics[width=\linewidth]{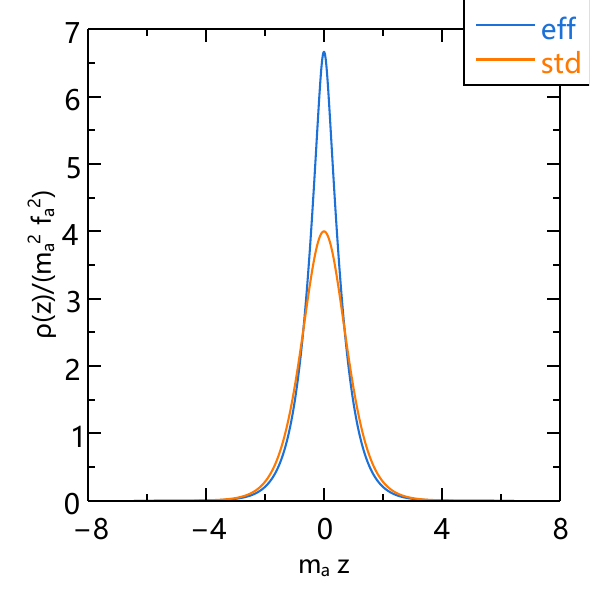}
		\end{minipage}
		\hfill
		\begin{minipage}[t]{0.3\textwidth}
			\includegraphics[width=\linewidth]{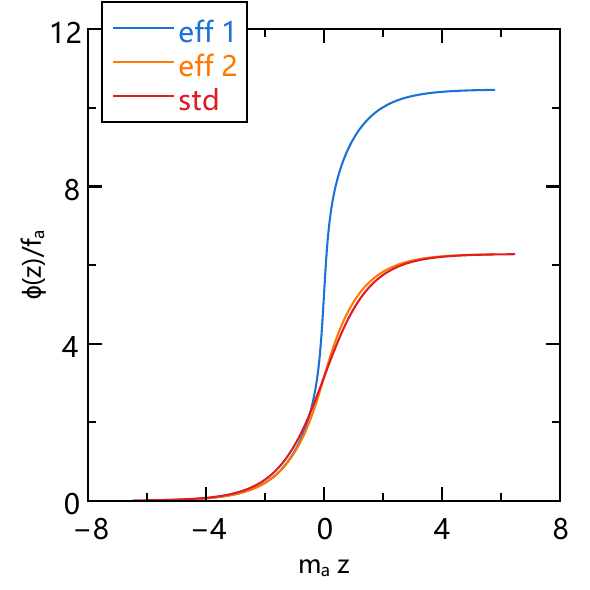}
		\end{minipage}
		\hfill
		\begin{minipage}[t]{0.3\textwidth}
			\includegraphics[width=\linewidth]{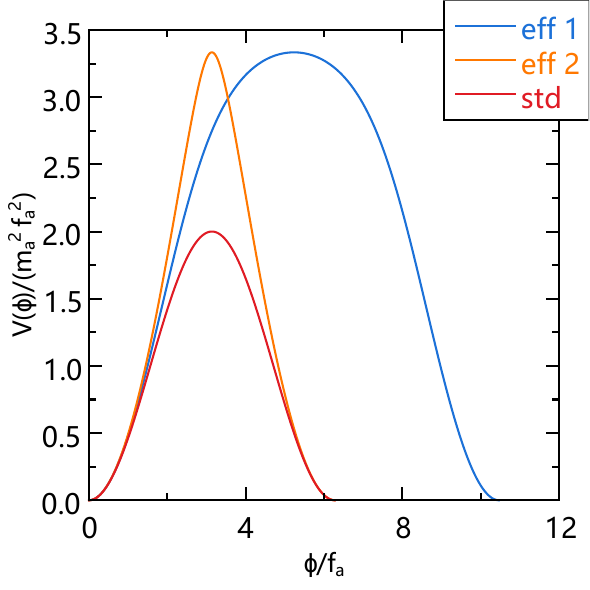}
		\end{minipage}
		
		\vspace{0.2cm}
		
		\begin{minipage}[t]{0.3\textwidth}
			\includegraphics[width=\linewidth]{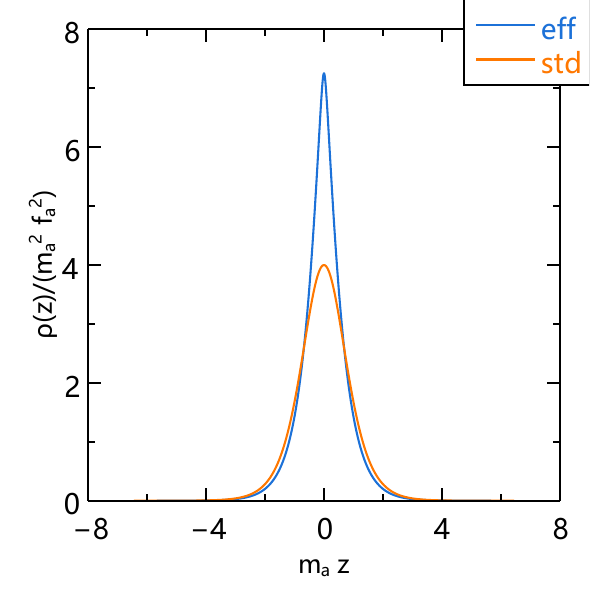}
		\end{minipage}
		\hfill
		\begin{minipage}[t]{0.3\textwidth}
			\includegraphics[width=\linewidth]{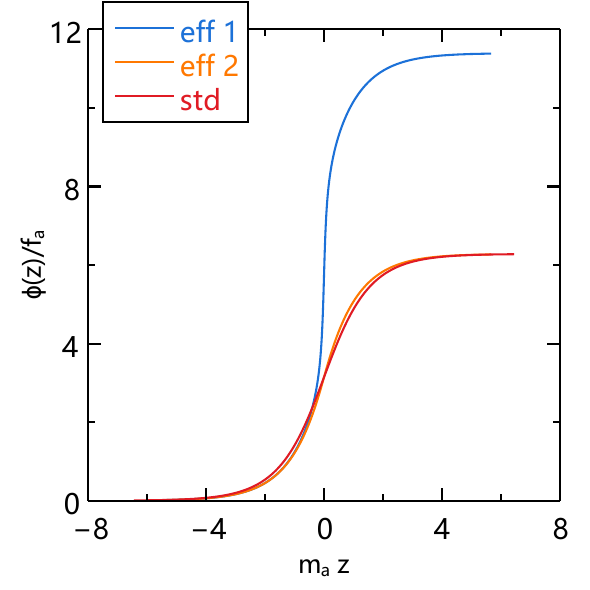}
		\end{minipage}
		\hfill
		\begin{minipage}[t]{0.3\textwidth}
			\includegraphics[width=\linewidth]{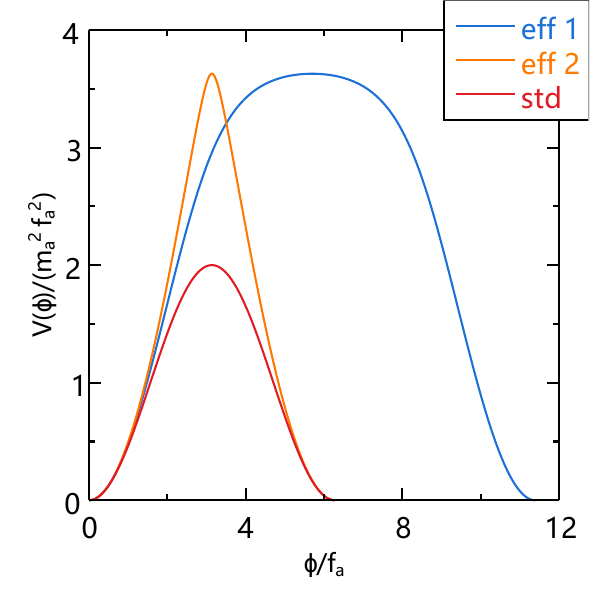}
		\end{minipage}
		\caption{In the left panels, the blue and orange curves represent the energy density distributions in the effective and standard single-phase models, respectively. In the middle and right panels, the blue and orange curves represent the DW configurations and scalar potentials in terms of the effective fields defined in Eq.~\eqref{heavy:phieff1} and Eq.~\eqref{heavy:phieff3}, respectively, while the red curves represent the DW configurations and scalar potentials in the standard single-phase model. The first, second, and third rows correspond to realizations (i), (ii), and (iii), respectively.}
		\label{fig:compare}
	\end{figure}

	\begin{itemize}
		\item[(i)] Hierarchy between the explicit breaking scales\\
		We assume $\Lambda_1\ll\Lambda_2$, so that
		\begin{equation}
			\frac{m_a^2}{m_b^2}\sim\xi\sim \zeta\sim\mathcal{O}\Big(\frac{\Lambda_1^4}{\Lambda_2^4}\Big). 
		\end{equation}
		The axion orbit can be approximated as
		\begin{equation}
			Y=\xi\sin X+2\pi m+\mathcal{O}(\xi^2),
		\end{equation}
		Then, the effective Lagrangian can be approximated as
		\begin{equation}
			\mathcal{L}_{\text{eff}}=\frac{1}{2}\partial_\mu \phi_{\text{eff}}\partial^\mu \phi_{\text{eff}}-\Lambda_1^4\Big(1-\cos \frac{\phi_{\text{eff}}}{f_{a}}\Big)+\mathcal{O}(\xi).
		\end{equation}
		In this scenario, the effective Lagrangian is consistent with the standard single-phase model at leading order.
		
		The overall scales of $(\Lambda_1^4,\Lambda_2^4)$ and $(f_1,f_2)$ can be absorbed into the scales of $m_{a}$ and $f_{a}$, leaving $\xi$, $\zeta$, $R_{\sigma}$, and $R_{\rho}$ invariant. We can therefore reduce the input parameters to $\Lambda_2^4/\Lambda^4_1$, $f_2/f_1$, and the integers $n_1,n_2,m_1,m_2$.
		
		As a simple example, we choose $\Lambda_2^4/\Lambda_1^4=100, f_2/f_1=1, n_1=m_1=m_2=1,n_2=2$. With these parameters, we obtain
		\begin{equation}
			\frac{m_a^2}{m_b^2}\approx0.0024,\quad \xi\approx-0.015,\quad \zeta\approx0.023,\quad R_{\sigma}\approx1.0074,\quad R_{\rho}\approx1.023.
		\end{equation}
		The corresponding energy density distributions, DW configurations, and scalar potentials are presented in the first row of Fig.~\ref{fig:compare}. In this figure, the energy density $\rho$ and the scalar potential $V$ are both divided by $m_a^2f_a^2$, while the fields $\phi_{\text{eff}}$ and $a$ are divided by $f_a$ to factor out the overall scales. The coordinate $z$ is also rescaled by a factor of $m_a$.
		In particular, in the middle panel, the red curve almost completely overlaps with the orange curve.
		
		These results imply that, in this realization, the differences between the effective and standard single-phase models are tiny.
		
		\item[(ii)] Hierarchy between the spontaneous breaking scales\\
		For simplicity, we assume $f_1\ll f_2$, so that
		\begin{equation}
			\frac{m_a^2}{m_b^2}\sim\mathcal{O}\Big(\frac{f^2_1}{f^2_2}\Big).
		\end{equation}
		Then,
		\begin{equation}
			\xi\simeq-\frac{\Lambda_1^4 n_1}{\Lambda_2^4 m_1},\quad \zeta\simeq\frac{\Lambda_1^4 n_1^2}{\Lambda_2^4 m_1^2},
		\end{equation} 
		where $|\xi|<1$ is required.
		As a simple example, we choose $\Lambda_2^4/\Lambda_1^4=1.5, f_2/f_1=100, n_1=m_1=m_2=1,n_2=2$. With these parameters, we obtain
		\begin{equation}
			\frac{m_a^2}{m_b^2}\approx2.4\times10^{-5},\quad \xi\approx-0.67,\quad \zeta\approx0.67,\quad R_{\sigma}\approx1.15,\quad R_{\rho}\approx1.67.
		\end{equation}
		The corresponding energy density distributions, DW configurations, and scalar potentials are presented in the second row of Fig.~\ref{fig:compare}. The left panel and $R_{\rho}$ show that the difference in the energy density distribution is mainly concentrated near the DW center. In the middle and right panels, the blue curve deviates significantly from the other two curves as the second vacuum is approached because the effective field used for the blue curve is defined by an expansion about the first vacuum. 
		
		\item[(iii)] Integer alignment\\
		For simplicity, we assume
		\begin{equation}
			|D|\ll|\operatorname{Max}(n_1^2,n_2^2,m_1^2,m_2^2)|\quad \text{and}\quad |n_1|\sim |n_2|\sim |m_1|\sim |m_2|,
		\end{equation}
		then
		\begin{equation}
			\frac{m_a^2}{m_b^2}=\mathcal{O}\Big(\frac{D^2}{n_1^4}\Big),\quad \xi\simeq-\frac{\Lambda_1^4 n_2}{\Lambda_2^4 m_2},\quad \zeta\simeq\frac{\Lambda_1^4 n_2^2}{\Lambda_2^4 m_2^2},
		\end{equation}
		where $|\xi|<1$ is required.
		As a simple example, we choose $\Lambda_2^4/\Lambda_1^4=1.5, f_2/f_1=1,n_1=11,n_2=m_1=10,m_2=9$. With these parameters, we obtain
		\begin{equation}
			\frac{m_a^2}{m_b^2}\approx 6.2\times 10^{-6},\quad \xi\approx-0.74,\quad \zeta\approx0.81,\quad R_{\sigma}\approx1.16,\quad R_{\rho}\approx1.81.
		\end{equation}
		Note that 
		\begin{equation}
			\frac{9}{11}-\zeta\approx0.0042,
		\end{equation}
		so this example is very close to the boundary of validity. The corresponding energy density distributions, DW configurations, and scalar potentials are presented in the last row of Fig.~\ref{fig:compare}. These results are similar to those of the second realization.
		
	\end{itemize}
	
	The numerical results above indicate the following: 
	\begin{itemize}
		\item[1.] If the mass hierarchy is realized by the hierarchy between the explicit breaking scales, the difference between the effective single-phase model and the standard one is suppressed by the hierarchy.
		\item[2.] If the mass hierarchy is realized by the hierarchy between the spontaneous breaking scales or by integer alignment, there are non-negligible differences between the effective single-phase model and the standard one. $R_{\sigma}$ can reach $\mathcal{O}(1.1)$ and $R_{\rho}$ can exceed $1.5$.
		\item[3.] The middle and right panels in Fig.~\ref{fig:compare} imply that the effective field defined in Eq.~\eqref{heavy:phieff3} is better suited to describing DW configurations than that defined in Eq.~\eqref{heavy:phieff1}.
	\end{itemize}
	\clearpage 

	\bibliographystyle{JHEP}
	\bibliography{Axion_cite.bib}
\end{document}